\documentclass{article}

\PassOptionsToPackage{numbers, compress}{natbib}
 
\usepackage[preprint]{neurips_data_2023}

\usepackage[utf8]{inputenc} 
\usepackage[T1]{fontenc}    
\usepackage{hyperref}       
\usepackage{url}            
\usepackage{booktabs}       
\usepackage{amsfonts}       
\usepackage{nicefrac}       
\usepackage{microtype}      
\usepackage{xcolor}         

\newcommand{\algo}{{\texttt{XFlow}~}}
\usepackage{enumitem,amssymb}
\newlist{todolist}{itemize}{2}
\setlist[todolist]{label=$\square$}
\setlist{leftmargin=10pt}
\usepackage{wrapfig}
\usepackage[font=small, format=plain, labelfont={bf}]{caption} 
\usepackage{graphicx}
\usepackage{amsmath}
\DeclareMathOperator*{\argmax}{arg\,max}
\DeclareMathOperator*{\argmin}{arg\,min}
\usepackage{subfigure}
\usepackage{listings}
\usepackage{lipsum}
\usepackage{multirow}

\title{\algo: Benchmarking Flow Behaviors over Graphs}

\author{%
Zijian Zhang$^*$, Zonghan Zhang\thanks{equal contribution}, Zhiqian Chen \\
  Department of Computer Science and Engineering \\
  Mississippi State University, Mississippi State, MS 39762\\
  \{zz242,zz239\}@msstate.edu, zchen@cse.msstate.edu \\
}

\begin{document}

\maketitle

\begin{abstract}
    The occurrence of diffusion on a graph is a prevalent and significant phenomenon, as evidenced by the spread of rumors, influenza-like viruses, smart grid failures, and similar events. Comprehending the behaviors of flow is a formidable task, due to the intricate interplay between the distribution of seeds that initiate flow propagation, the propagation model, and the topology of the graph. 
    The study of networks encompasses a diverse range of academic disciplines, including mathematics, physics, social science, and computer science. This interdisciplinary nature of network research is characterized by a high degree of specialization and compartmentalization, and the cooperation facilitated by them is inadequate. From a machine learning standpoint, there is a deficiency in a cohesive platform for assessing algorithms across various domains.
    One of the primary obstacles to current research in this field is the absence of a comprehensive curated benchmark suite to study the flow behaviors under network scenarios.
    To address this disparity, we propose the implementation of a novel benchmark suite that encompasses a variety of tasks, baseline models, graph datasets, and evaluation tools.
    In addition, we present a comprehensive analytical framework that offers a generalized approach to numerous flow-related tasks across diverse domains, serving as a blueprint and roadmap. 
    Drawing upon the outcomes of our empirical investigation, we analyze the advantages and disadvantages of current foundational models, and we underscore potential avenues for further study.
    The datasets, code, and baseline models have been made available for the public at: 
    \url{https://github.com/XGraphing/XFlow}.
\end{abstract}

\section{Introduction}

Graphs provide a powerful representation of complex systems, such as social networks \cite{knoke2019social}, epidemic spreading network \cite{li2021modeling}, financial transaction network \cite{gao2020analysis}, power network \cite{boyaci2021graph} and brain networks \cite{lynn2019physics}. 
The functionality of a graph is largely controlled by its topology, flow dynamics, and the relationship between them. 
This paper discusses the regulation of flow dynamics, referred to as graph flow in this paper, through the structure of a network. As an illustration, data is transmitted amidst individuals who follow each other on social media platforms, contagion disseminates amidst close acquaintances, the distribution of power is regulated by switches situated along power lines, and blood and electrical impulses traverse the neural network of the brain. 
Conversely, the occurrence of flows can lead to adverse consequences, thereby requiring a structural alteration to avert the collapse of the system. 
Instances of managing power overload by flipping switches, halting seizures by removing brain tissue, preventing the spread of an epidemic through partial lockdowns, and restricting the spread of disinformation by disconnecting crucial nodes on social networks can be observed.

The academic discipline of graph flow, which pertains to the study of flow over graphs, has attracted significant interest from a wide range of scholarly fields, such as mathematics, physics, social science, and computer science. 
Nevertheless, these domains frequently operate independently, resulting in a significant disparity between them. Consequently, the absence of a unified platform has resulted in the fragmentation of network research, preventing convergence toward a cohesive community.
Specifically, there are several domains doing relevant research topics in different sciences.
In machine learning, research focuses on node-, link-, and graph-level attribute prediction \cite{wu2020comprehensive,zhou2020graph}, and also explores a few emerging topics such as generative graphs \cite{guo2022systematic}.
The concept of network flow pertains to the field of mathematical graph theory and involves the optimization of routes to facilitate the movement of resources or quantities across a graph ("graph" and "network" will be utilized interchangeably throughout this paper) \cite{ahuja_network_1993,cormen_introduction_2009,edmonds_theoretical_1972,ford_maximal_1956}. 
For example, in transportation, network flow can be used to maximize the flow of goods through a supply chain, determining the most efficient routes and quantities to transport between different locations. 
Apart from route optimization, the source nodes hold a crucial position in enhancing flow, which is the primary goal of influence maximization. This involves identifying a cluster of individuals within a network who possess substantial influence and can effectively facilitate the spread of information, ideas, or initiatives \cite{domingos_mining_2001,kempe_maximizing_2003,cohen_efficient_2003,leskovec_cost-effective_2007}. 
The field of network science has recently been investigating novel frameworks, such as higher-order networks \cite{Boccaletti2023May} and multilayer networks \cite{boccaletti2014structure}. 
Distinctive characteristics are inherent to each application domain from a domain-specific perspective. For instance, non-replicable flow, such as power and brain flow, and replicable flow, such as disease and rumor, are some of the unique features that can be observed in different application domains.

Due to the widely-recognized importance across various domains, a number of datasets and benchmarks have been developed for network and graph research, including
\text{graph data collections}~\cite{hu2020open,nr-aaai15,snapnets},
\text{graph learning}~\cite{Fey/Lenssen/2019,wang2019deep,JMLR:v22:21-0343,rozemberczki2021pytorch,tp3d,8547527,landry2023xgi}, and
\text{diffusion model}~\cite{ndlib,McCabe2022joss,jenness2018epimodel} (see Section \ref{sec:related_work} for details).
However, none of them provides a comprehensive benchmark suite for flow problems on graphs.
In order to facilitate foundational investigations into graph flow and address existing challenges within the scholarly community,
The main aim of this open-source initiative is to create a platform that facilitates the development of a computational framework suitable for gaining a comprehensive understanding of the complex behaviors demonstrated by dynamics over networks during their evolutionary phases. This encompasses activities such as tracing the trajectory of dynamics, identifying their origins, and determining the most effective control strategies. In order to commence this dedicated undertaking, we are inaugurating a benchmark project called \algo, which will be open-source and serve as the initial phase of our overarching mission.

Our work has the following highlights:
\textbf{(1) Offer a unified framework:} The proposed unified framework seeks to present a holistic methodology for addressing the diverse range of related works, with the potential to serve as a model for tackling the core concerns surrounding graph flows.
\textbf{(2) Provide a flexible development environment}: The current study introduces a versatile and flexible framework that is easily accessible and can be promptly utilized and expanded upon. This framework includes a variety of pre-designed tasks that encompass a diverse range of graph configurations, diffusion models, and seed distributions.. 


\textbf{Broader Impacts:}
The objective of this study is to produce significant findings in the area of network analysis, thereby contributing to the fundamental scientific understanding required for analyzing, controlling, and improving behaviors in networked environments.  The undertaking bears substantial potential to influence the domain of computer science research by emphasizing the significance of comprehensive modeling of networked systems, utilizing insights from diverse fields such as mathematics, physics, computer science, social science, and network science.
The utilization of data-driven methodologies facilitated by recent progressions in machine learning holds promise in mitigating the aforementioned difficulties. Simultaneously, the integration of innovative techniques derived from other domains of science has the potential to propel progress in the field of machine learning.
The convergence of mathematics, physics, network science, social science, and machine learning through collaborative efforts can result in the emergence of various advantages in the formulation of methodologies.
The proposed framework demonstrates versatility across a broad range of applications, including but not limited to public health, infrastructure security, social network information propagation, viral marketing, epidemiology, transportation, and power networks. This versatility makes it a valuable tool for practitioners in these fields.

\textit{Limitations and Potential Negative Societal Impacts:} 
Our project aims at the machine learning community, which may not be friendly to other domains or the non-python communities.
An adverse algorithm can utilize this platform and can be developed for misinformation spreading.

\section{Related Work}
\label{sec:related_work}

\textbf{Existing Benchmarks and Datasets.}
There exists a number of open-source datasets and tools that are often used by network applications:
(1) \textit{Graph data repositories.}
Open Graph Benchmark (OGB) is a comprehensive benchmark designed to evaluate and advance the performance of graph representation learning algorithms on real-world graph data. \cite{hu2020open}.
The Network Repository is an online resource that hosts a vast collection of network datasets, including over 800 network datasets across various domains, including social networks, biological networks, transportation networks, and more \cite{nr-aaai15}.
The Stanford Large Network Dataset Collection (SNAP) is a comprehensive repository that provides a wide range of real-world network datasets for research purposes \cite{snapnets}.
(2) \textit{Graph learning.}
PyTorch Geometric is a powerful library for handling and analyzing geometric deep-learning tasks in PyTorch. With a rich set of functionalities and pre-built modules, it enables researchers and practitioners to easily work with graph-structured data, apply graph neural networks, and develop models for various graph-based applications \cite{Fey/Lenssen/2019}.
Another popular graph learning library termed DGL (Deep Graph Library), is a flexible and efficient library for graph neural network (GNN) research and development \cite{wang2019deep}.
DIVE (Dive into Graphs) is an interactive platform that allows users to explore and analyze graph data through a visual interface. With its user-friendly tools and intuitive visualizations, DIVE enables users to gain insights, perform graph-based analytics, and discover patterns within complex network structures \cite{JMLR:v22:21-0343}.
PyTorch Geometric Temporal is an extension of PyTorch Geometric that focuses on spatiotemporal graph analysis \cite{rozemberczki2021pytorch}.
Torch-points3d is a versatile library built on PyTorch Geometric, specifically designed for processing and analyzing 3D point cloud data, providing a comprehensive suite of tools for deep learning tasks in 3D computer vision \cite{tp3d}.
GraphChallenge.org is an online platform that hosts a series of graph analytics competitions, aiming to foster advancements in graph processing techniques and benchmark the performance of state-of-the-art graph algorithms on large-scale graph datasets \cite{8547527}
The XGI library is a resource that offers a range of data structures and algorithms that can be utilized for the purpose of modeling and analyzing complex systems that involve group interactions of a higher-order nature \cite{landry2023xgi}.
(3) \textit{Diffusion model.}
NDlib is a powerful Python library that provides a comprehensive set of tools for simulating and analyzing the dynamics of diffusion processes on complex networks \cite{ndlib}.
Cosasi is a Python library that facilitates graph diffusion source inference, which is the process of identifying the origin or source of information propagation in a network \cite{McCabe2022joss}.
EpiModel is a versatile R package that offers a comprehensive framework for simulating and analyzing the spread of infectious diseases on dynamic social networks, providing researchers with valuable tools for studying and predicting the dynamics of epidemics \cite{jenness2018epimodel}.

\textbf{Graph Flow Problems.}
\textit{Influence Maximization (IM)} is the process of finding a set of initial seeds to spread the influence on the network as much as possible. The problem has been proven to be NP-hard~\cite{kempe_maximizing_2003}, thus is often approached by approximations. As the first approximation attempt, Kempe et al.~\cite{kempe_maximizing_2003} proposed a simulation-based greedy algorithm, which achieves an approximation ratio of $1-\frac{1}{e}$ but is too time-consuming to be applied to large networks. To reduce the time complexity, a thread of simulation-based methods is developed~\cite{leskovec2007cost, goyal2011celf++,chen2013information}. Although great effort has been made to accelerate the process of the simulation-based methods, the complexity is still unacceptably high for the enormous online social networks~\cite{arora2017debunking, tang2014influence}. 
To avoid the computational burden of simulations, researchers try to estimate the spreading power of each node by certain proxies. The proxy-based methods started from simple heuristic measures such as degree, PageRank~\cite{page1999pagerank}, and eigen-centrality~\cite{zhong2018identifying}. They later turned to influence-aware and diffusion model-aware proxies~\cite{chen2009efficient, kimura2009blocking,chen2010scalable,chen2010scalableLT,tong2010vulnerability,goyal2011simpath,yan2019minimizing,zhang2022blocking} to obtain better estimations of the influence spread brought by the seeds.
\textit{Influence blocking maximization (IBM)} \cite{budak2011limiting, he2012influence,ibrahim2018controlling} aims to develop feasible strategies to minimize unwanted influence spreads on the network. This problem draws attention due to the significantly negative consequence brought by negative influences such as rumors, fake news \cite{fan2013least, allcott2017social,chowdhury2020joint} and infectious diseases \cite{ozili2020spillover, sher2020impact, area2017ebola}.
The common approach to blocking influences, node removal, is identified as a major strategy for limiting the diffusion of unwanted information~\cite{tong2010vulnerability,wang2013negative,shi2019adaptive}. Node removal has also been utilized in the real-world application such as vaccination, which is one of the most effective ways to prevent infectious diseases~\cite{medlock2009optimizing,clem2011fundamentals, glass2006targeted}, and user blocking on social networks~\cite{kaur2017blocking,yan2019minimizing} to stop rumors from spreading. 
Like IM, IBM can also be modeled as a combinatorial optimization problem that is NP-hard. Thus, the majority of current methods follow a greedy pattern \cite{wang2013negative,yan2019minimizing,ma2016identifying,tong2010vulnerability,kimura2009blocking,tong2012gelling,budak2011limiting,tong2017efficient,ibrahim2018controlling} to reach an approximation solution. 
\textit{Source Localization (SL)} of information diffusion is the process of inferring the initial diffusion sources given the current diffused observation. This problem has many applications, such as rumor source localization and flaw detection in power grids~\cite{jiang2016identifying, shelke2019source}. Various methods~\cite{dong2019multiple, prakash2012spotting, wang2017multiple, zhu2017catch, zhu2014information} have been proposed by scholars to identify sources of an observed diffusion. Early approaches~\cite{prakash2012spotting} identified a single source of an infectious disease under the Susceptible-Infected (SI) diffusion model utilizing the Minimum Description Length principle. 
Methods~\cite{zang2015locating, zhu2017catch, zhu2014information} have been proposed to transfer the source localization problem to predict rumor sources under the Susceptible-Infected-Recovered (SIR) diffusion model with only partial observation. 
Wang et al. \cite{wang2017multiple} proposed a model named LPSI,  which can detect multiple diffusion sources without information on the diffusion models. Except for LPSI, existing methods are usually designed for specific diffusion models (e.g., SI, IC, or LT). Even LPSI assumes the seeds are included in the activated subgraph, hinting that the diffusion model cannot be SIR.
\textit{Spatial-Temporal Graph Neural Networks} are a class of deep learning models that corporates spatial and temporal information, and predict the dynamic processes occurring in complex networks, such as traffic flow, social interactions, and environmental phenomena \cite{jiang2022graph,jin2023spatio}. Most of these applications utilize black-box neural network configurations, which may not provide transparent insights into their functionality.

\section{Tasks, Datasets and \algo}

\subsection{Overview}

Despite the close relationship among graph flow problems, they have been previously studied separately as individual tasks by researchers. Consequently, there is a lack of a unified framework that can conclude these problems. To address this deficiency, we propose a general conceptual framework that includes the key factors involved in graph flow studies. 
The proposed framework includes five factors: source node set $\Omega$, activated sub-graph $\sigma$, graph structure $G$, diffusion model $D$, and flow routing $\mathcal{F}$. A graph flow problem designates some factors as the input while some other factors act as the output, and thereby it is possible to deduce the remaining factors based on the partial factors.

\begin{wrapfigure}[12]{r}[0pt]{0.5\linewidth}
    \centering
    \includegraphics[width=1\linewidth]{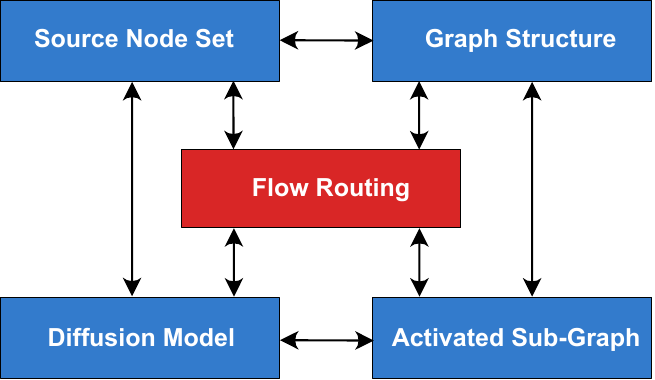}
    \caption{The framework of graph flow problems.}
    \label{fig:framework}
\end{wrapfigure}

Current graph flow problems can be conceptualized using our proposed framework. As depicted in Fig.~\ref{fig:framework}, the five factors represented constitute the majority of existing graph flow problems. 
The diffusion model frequently serves as a moderator, and the interdependence among the other remaining factors is dependent on the diffusion model. Nevertheless, it maintains a consistent value within a specified problem context and should generally not be modified. Moreover, it has not been classified as a result of graph flow issues. So the diffusion model may also serve as a noteworthy output of interest. 
By considering the details pertaining to the source, activation, and graph structure, it is possible to acquire a diffusion model that effectively captures the association and accurately portrays the diffusion trend.
The significance of flow routing, which accounts for the dynamics involved in the propagation process, has been widely disregarded in academic literature. This is primarily attributed to the complexity of the process that occurs between the diffusion source and the propagation outcome. The identification of the flow pattern between these two factors bears significant practical importance. Prospective studies that seek to determine the routing of information flow based on the seed set and activated subgraph are anticipated and hold great potential.

\subsection{Tasks}

\begin{table}[htpb!]
\centering
\setlength{\fboxsep}{1pt} 
\fbox{
\scalebox{0.84}{
\begin{tabular}{c|c|c|c}
\textbf{Flow Tasks} & \textbf{Explicit Input}             & \textbf{Implicit Input}                      & \textbf{Output}     \\
\hline
IM         &   graph structure $G$    &  diffusion model $D$    & seed set $\Omega$               \\
\hline
IBM        & graph structure $G$, seed set $\Omega$    &  diffusion model $D$      & graph structure $\Delta G$               \\
\hline
SL         & observation $\sigma$   &  graph structure $G$, diffusion model $D$    & seed set $\Omega$                         \\
\hline
Diffusion Pattern         & seed set $\Omega$, observation $\sigma$   &  graph structure $G$, flow routing $\mathcal{F}$   & diffusion model $D$                         \\
\hline
Path Study         & seed set $\Omega$, observation $\sigma$   &  graph structure $G$, diffusion model $D$    & flow routing $\mathcal{F}$                       \\
\end{tabular}
}
}
\caption{Three common flow tasks (top three) and two potential tasks.}
\label{tab:flowtasks}
\end{table}

Three tasks were implemented to demonstrate our proposed framework, including influence maximization (IM), influence blocking maximization (IBM), and source localization (SL). They are formally identified below according to the framework presented in Figure ~\ref{fig:framework}. Table~\ref{tab:flowtasks} summarizes those three and some possible future tasks that can be included in the framework, and detailed formulations are elaborated below.

Given the graph structure information $G = (V, E, A)$ as the input and a pre-defined seed budget $k\in \mathbb{N}^{+}$, IM methods attempt to optimize the selection of a $k$-sized source node set $\Omega$ to maximize the expected influence spread $|\sigma(\Omega)|$ which is the size of the final activated subgraph $\sigma(\Omega)$. Except for the heuristic methods such as PageRank, degree, and eigen-centrality, the simulation-based and proxy-based methods also require a known diffusion model $D$ such as IC, LT, SI, or SIR. Thus, when the budget requirement is met ($|\Omega| = k$), IM aims at finding a seed set $\Omega$ such that:
\begin{equation}\small
    \Omega = \argmax_{\Omega^*}|\sigma(\Omega^*|G, D)|.
\end{equation}

IBM problem can be either against a set of known diffusion seeds or to prevent any future diffusion on the network. On the one hand, when graph structure $G = (V, E, A)$, diffusion model $D$, and the seed set $\Omega$ are known inputs, an IBM method aims to find part of $G$, namely $\Delta G$ such that the expected influence spread $|\sigma(\Omega)|$ is minimized if $\Delta G$ is blocked before the diffusion starts. Note that $\Delta G$ can be a set of nodes or edges, and there is usually a budget constraining the size of $\Delta G$. On the other hand, if the diffusion seed set is unknown, the goal of an IBM method is to minimize the expected influence spread of any diffusion starting from any set of seeds. This is usually achieved by removing $\Delta G$ from $G$ to weaken its connectivity. Annotating $G_{new} = G - \Delta G$, we have
\begin{equation}\small
    \Delta G = \argmin_{\Delta G^*}|\sigma(\Omega|G_{new}, D)|
\end{equation}

Given a snapshot of graph $G = (V, E, A)$ as the observation of the diffusion status at a certain time, the activated subgraph $\sigma$ of the snapshot is the propagation result of an unknown seed set $\Omega$. The goal of an SL method is to find out $\Omega$ based on the activated subgraph $\sigma$, the graph structure $G$, and the diffusion model $D$. It is achieved by finding the most likely seed set resulting in the activated subgraph $\sigma$.
\begin{equation}\small
    \Omega = \argmax_{\Omega^*}Pr(\Omega^*|\sigma, G, D)
\end{equation}
\textbf{Future Expansion.} Our framework encompasses traditional flow tasks and can also accommodate emerging tasks, such as influence maximization on hyper-graphs or multi-layer graphs. 
As the graph structure varies across categories, the relationships among the seed set, the activated subgraph, and the flow routing vary as well. When the complete observations of a diffusion process are available, they can be employed to investigate the diffusion pattern, including how nodes are activated and how the influence is transmitted among nodes. Additionally, as previously mentioned, research on flow routing $\mathcal{F}$ is currently limited. It is expected that in forthcoming times, scholars will be motivated to unveil a comprehensible correlation between diffusion seeds and outcomes. Analyzing the flow routing is an inevitable measure to accomplish this objective. Given the seed set and the activated subgraph, the propagation path of the influence can be inferred and further facilitate influence control, as referred to in Table~\ref{tab:flowtasks} as Path Study. Another potential flow routing research topic is how a smart flow can adjust itself during propagation to achieve optimal results. Compared to the traditional black box model, this approach can more precisely model the dynamic decision-making of self-conscious influences, such as rumors and advertisements.
We also acknowledge the existence of a relationship between replicable and non-replicable flows. Existing graph flow tasks focus on replicable flows such as information and viruses. The non-replicable flows, such as traffic and logistics, exhibit both similarities and dissimilarities compared to the replicable flows. We hope that a general framework capable of incorporating both will be developed in the future.

\subsection{Implementation Configuration}

\textbf{Diffusion Models.}
Various diffusion models employ distinct mechanisms to depict the process by which a vertex is stimulated from an inactive state, under the influence of its neighboring vertices.
The independent cascade (IC) \cite{goldenberg_talk_2001} and linear threshold (LT) \cite{granovetter_threshold_1978} models are the most commonly used and well-studied progressive models. Well-known non-progressive models include the susceptible-infected (SI) model, the susceptible-infected-removed (SIR) model, and susceptible-infected-susceptible (SIS) models\cite{kermack_contribution_1927}. The ongoing stage of \algo will encompass the evaluation of IC, LT, and SI as the principal diffusion models, with intentions to integrate supplementary models in subsequent stages as expounded below.
There also exist many variants of SIR models, including
SEIR extends the basic SIR model by introducing an additional compartment called "Exposed" (E) \cite{Kermack1927};
SIRS where individuals can transition back to the susceptible state after recovering from the infection \cite{Anderson1980};
SIRD model expands upon the SIR model by including a compartment for "Dead" (D) individuals \cite{Hethcote2000};
SEIRS model combines the features of the SEIR and SIRS models by including an exposed compartment and allowing individuals to become susceptible again after recovery \cite{Heesterbeek1996}.
It should be noted that most current diffusion models fall under the category of mean-field style, wherein each entity exhibits identical diffusion behavior. Nevertheless, the personalized diffusion model provides a more comprehensive representation of real-life interactions, yet it is frequently disregarded \cite{guo2013personalized, li2015real, nguyen2016cost, chen2016real, li2017discovering, tian2020deep}.




\textbf{Graph Datasets.}
\algo incorporates existing dataset collections and can be extended easily.
Specifically, our implementation is compatible with any graph in NetworkX graph object \cite{hagberg2008exploring}, including graphs such as Watts-Strogatz small-world graphs, Barabási–Albert (BA), Erdős–Rényi (ER). See the full list in \texttt{Graph Generators} in NetworkX; 
Or graph objects in PyTorch Geometric such as Cora, Cite Seer, PubMed and co-purchasing networks (Amazon Photo, Computers) as well as synthetic graph generators. See \texttt{torch\_geometric.datasets} in PyTorch Geometric.

\textbf{Baseline.}
IM algorithms are organized into three main categories
(1) simulation-based methods, exemplified by algorithms like greedy~\cite{kempe_maximizing_2003}, CELF~\cite{leskovec2007cost}, and CELF++~\cite{goyal2011celf++}, apply Monte Carlo simulation to estimate the influence spread. This category's strength lies in its model generality, meaning that these algorithms can adapt to a multitude of diffusion models. Nevertheless, they tend to have lower computational efficiency and higher time complexity, which can be a drawback.
(2) proxy-based algorithms, such as pi~\cite{zhang2022blocking}, sigma~\cite{yan2019minimizing}, degree, and eigen-centrality, utilize proxy models to approximate the influence function, thereby tackling the NP-hard problem. These algorithms provide the advantage of high practical efficiency. However, they come with the caveat of not having theoretical guarantees, which might limit their reliability in certain scenarios.
(3) sketch-based IM algorithms, for example, SKIM~\cite{cohen2014sketch}, IMM~\cite{tang2015influence}, and RIS~\cite{borgs2014maximizing}, use compact data structures called sketches to approximate the influence spread in a network. It focuses on improving the efficiency of the simulation theoretically and reducing the time complexity while preserving a constant approximation guarantee. However, they might not be as general in using diffusion models as simulation-based algorithms and have less practical efficiency than proxy-based algorithms.
Like IM, the state-of-the-art method for IBM is the simulation-based greedy algorithm. Given a known seed set, the node that reduces the influence spread evaluated by simulations is selected in each iteration until the budget is used. The other IBM methods implemented include degree, eigen-centrality, pi, and sigma. The proxy-based methods, such as pi and sigma, are customized for IM tasks. Thus, their performance on the IBM task cannot be guaranteed.
\begin{figure}[htpb!]
    \centering
    \subfigure{\includegraphics[width=0.3\textwidth]{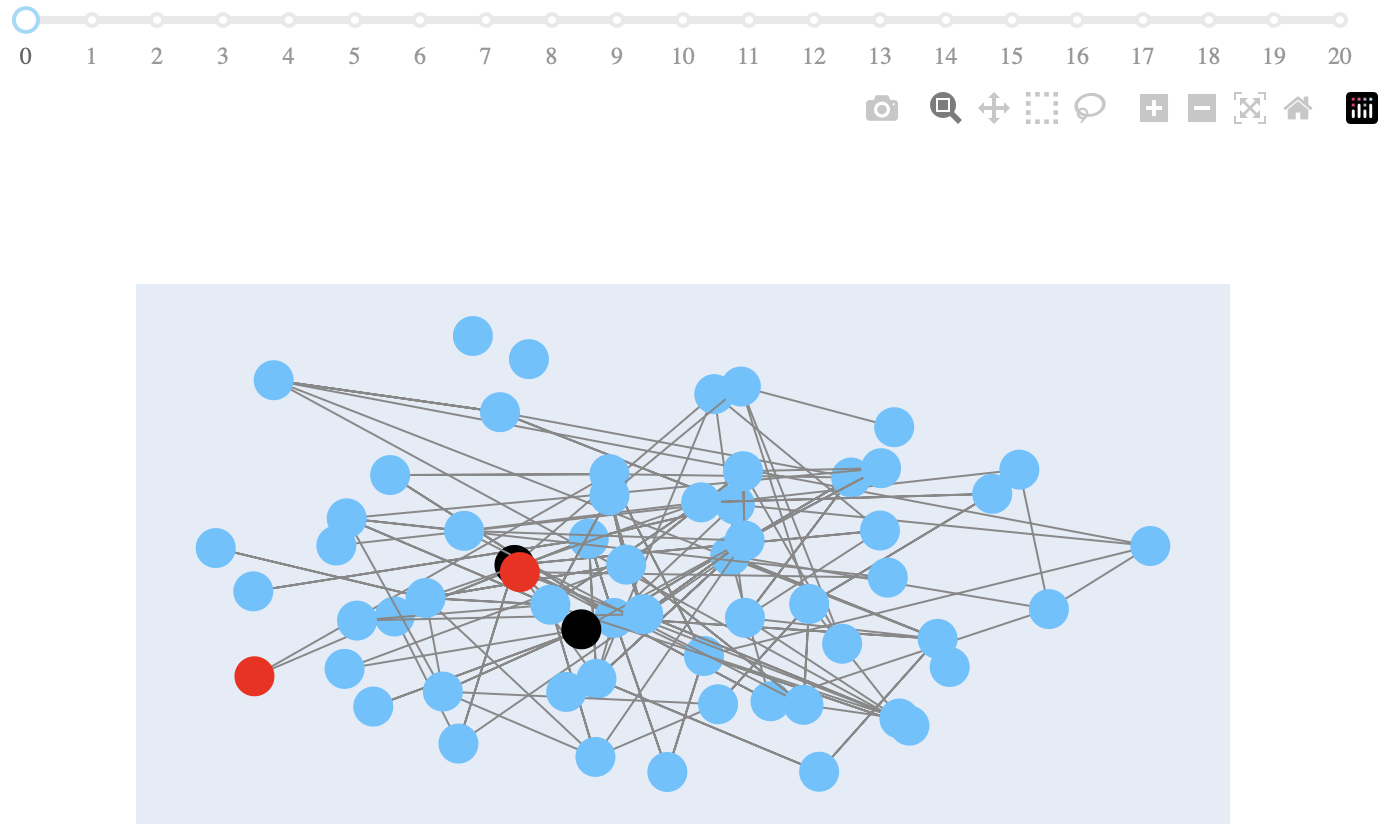}}
    \subfigure{\includegraphics[width=0.3\textwidth]{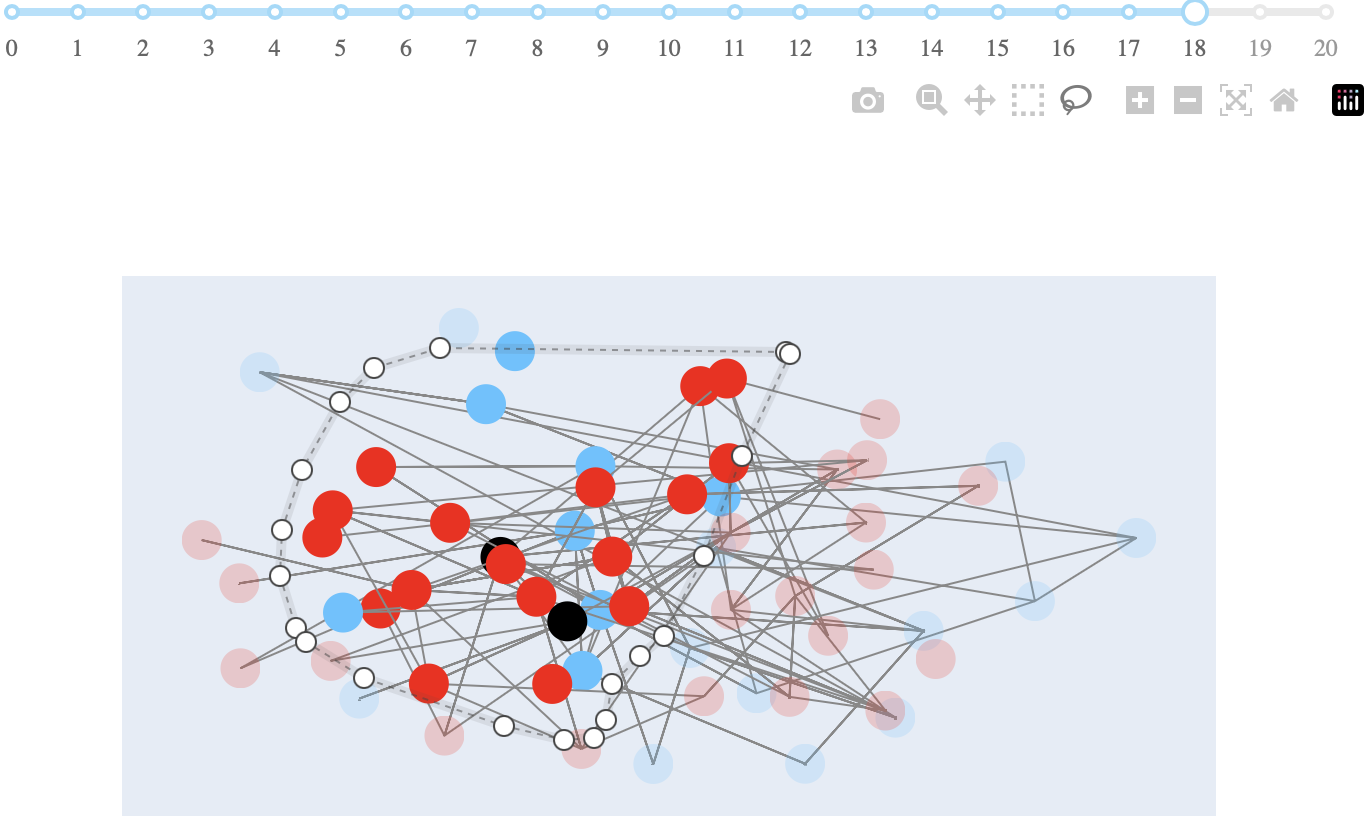}}
    \subfigure{\includegraphics[width=0.3\textwidth]{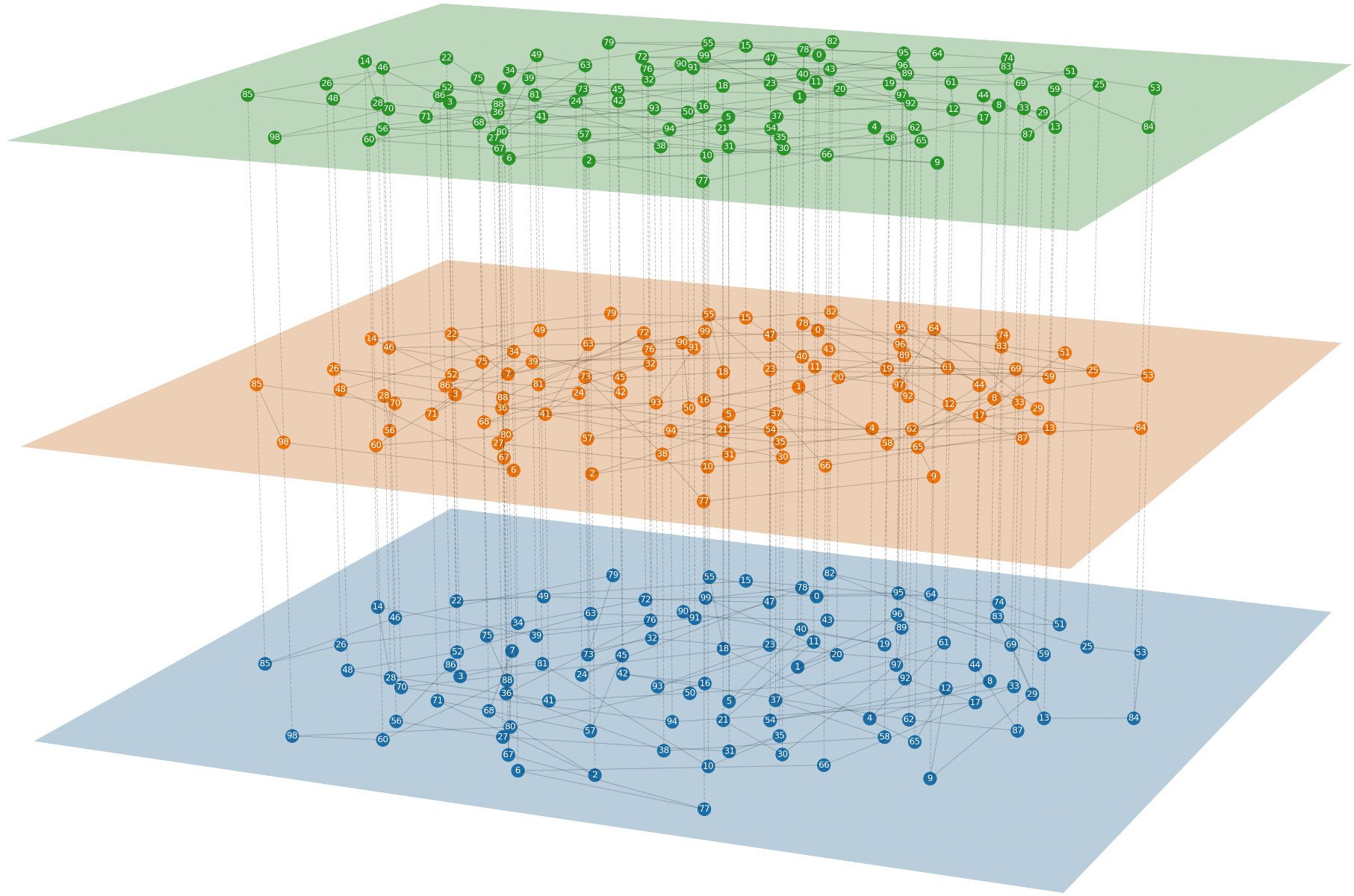}}
    \caption{Visualization: (a) spreading at step 0, (2) step 18 with selected nodes for inspection, (3) multilayer}
    \label{fig:vis}
\end{figure}
SL starts from single-source methods and later turns to multi-source methods. 
Based on the different types of observations available, single-source network observation-based methods are divided into complete observation, snapshot observation, and monitor observation. They employ different observational strategies to trace the information source. For complete observation, methods such as rumor centrality and eigenvector centrality come into play.
Snapshot observation methods like Jordan Centrality utilize network snapshots at specific intervals to locate the source. 
Multi-source methods include partition-based and ranking-based.
Partition-based algorithms divide the activated nodes into sections and attempt to identify the source within each partition. Ranking-based methods, such as NETSLEUTH~\cite{6413787}, rank nodes based on certain criteria to identify the source(s).

\begin{wrapfigure}[14]{r}[0pt]{0.45\linewidth}
\begin{lstlisting}
from xflow.dataset import cora, random, ba
from xflow.diffusion import ic, si
from xflow.seed import random, degree, eigen
from xflow.method.im import celf, sigma

# graphs to test
gs = [cora, random, ba]
# diffusion models to test
df = [ic, si]
# seed configurations to test
se = [random, degree, eigen]
# methods to test
me = [celf, sigma, imrank]
# configurations of experiments
rt = run(graph=gs, diffusion=df, seed=se, method=me, eval='im', epoch=10, output=['animation', 'csv', 'fig'])
\end{lstlisting}
\caption{API use example for IM}
\label{code:ready_to_use}
\end{wrapfigure}

\textbf{XFlow.}
The \algo library represents a Python implementation of our sophisticated benchmarking framework, designed to offer researchers and practitioners a cohesive and user-friendly toolset for conducting benchmark experiments effortlessly. With its inherent flexibility for extension and integrated visualization capabilities, the \algo library streamlines the development process.
As illustrated in the example showcased in Figure \ref{code:ready_to_use}, users can easily select specific graphs, diffusion models, and seed configurations through a streamlined interface. By employing the versatile \texttt{run} function, these chosen combinations can be executed in a batch fashion. In the aforementioned example, a total of 36 IM experiments will be conducted by combining 2 graphs, 2 diffusion models, 3 seed configurations, and 3 baseline methods. Each individual experiment will be repeated 10 times, providing essential statistical outputs such as the size of the activated subgraph. 
It is worth noting that certain input variables possess the capability to be furnished with parameters. For instance, the IC model can be configured to exhibit a probability of 30\% through the utilization of the parameter \texttt{ic(0.3)}. Similarly, the number of steps can be determined by incorporating the parameter \texttt{step=10} within the \texttt{run} function.
Furthermore, the library allows users to customize the output format, enabling dynamic animations in web format, the creation of intermediate CSV data for future use, as well as the generation of informative figures as exemplified in Figure \ref{code:ready_to_use}.
An illustrative example visualization is depicted in Figure \ref{fig:vis}, wherein we present a web-based format enabling dynamic time manipulation through a slider mechanism, accompanied by a zoom functionality that facilitates detailed scrutiny of node attributes.

\section{Experiments}









\textbf{IM \& IBM.}
We conducted experiments with three different settings to demonstrate the implementation of~\algo on IM and IBM. The first holds the diffusion model as SI with a beta value of 0.1 and the graph as the WS small-world graph sized 1000. The budget of IM and IBM algorithms varies from 5 to 30 in increments of 5. This setting is to reveal how the baselines perform when the budget changes. The second experiment fixes the budget and the graph while adjusting the beta of the SI model incrementally from 0.1 to 0.5, demonstrating the capability of~\algo to evaluate the methods with different diffusion models. Lastly, the diffusion model and the budget are stable across experiment setting 3. The size of the WS small-world graph changes from 200 to 1000 in increments of 200; this aims to demonstrate the algorithm's evaluation capabilities across different graph structures.
\begin{wrapfigure}[45]{r}[0pt]{0.48\linewidth}
    \centering
    \subfigure{\includegraphics[width=0.48\textwidth]{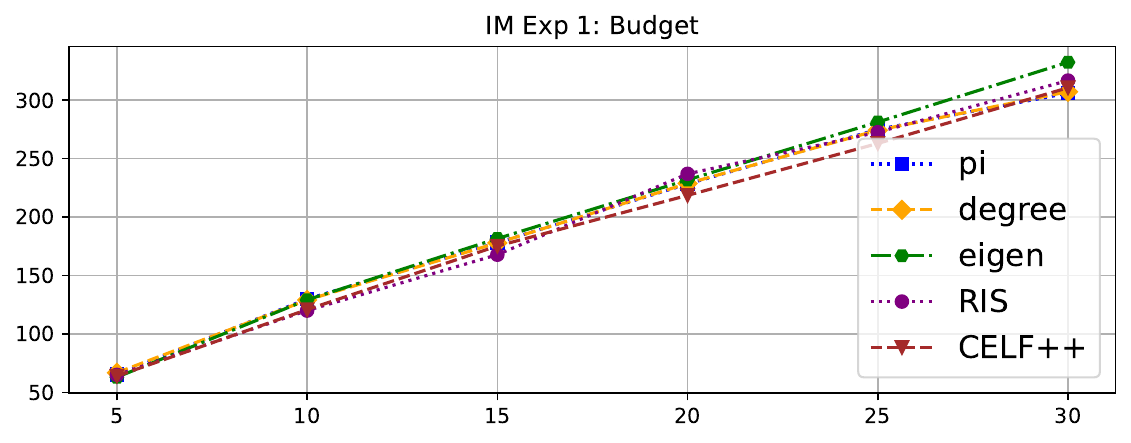}}
    \subfigure{\includegraphics[width=0.48\textwidth]{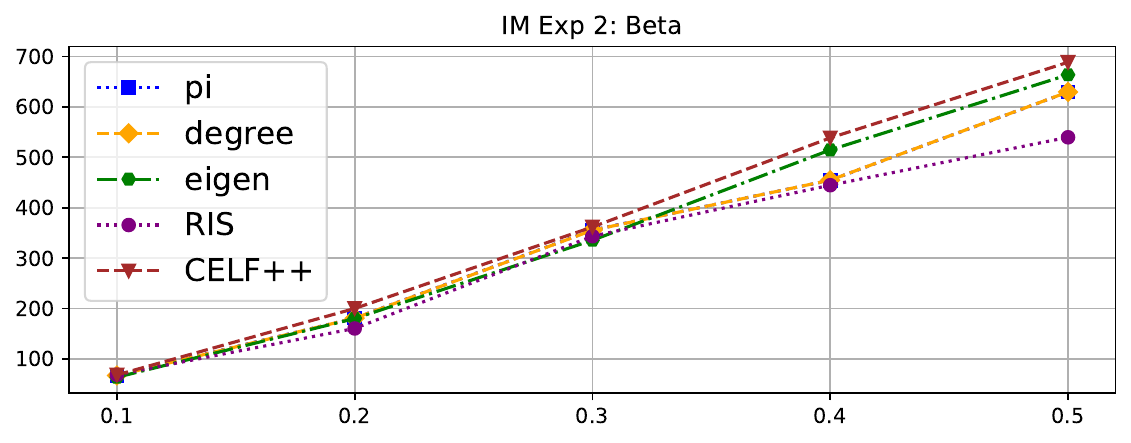}}
    \subfigure{\includegraphics[width=0.48\textwidth]{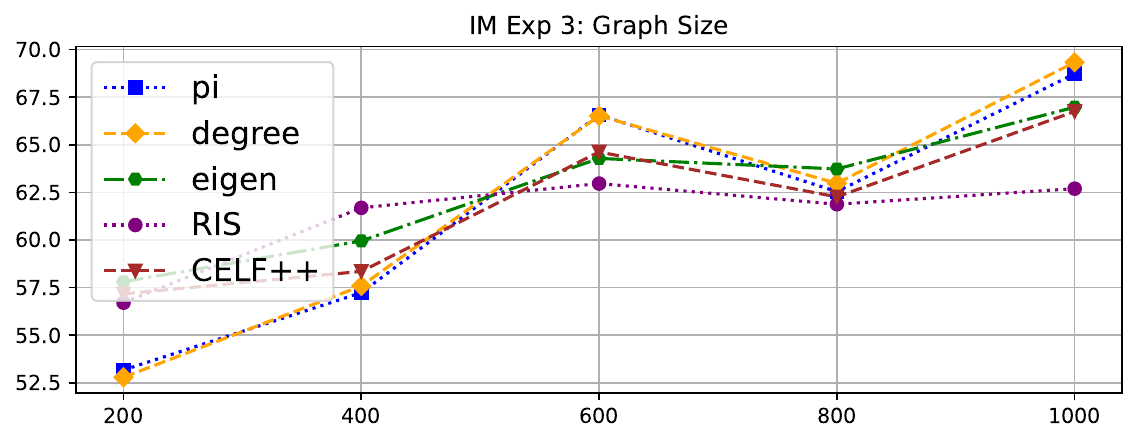}}
    \caption{Influence spread regarding (a) budget, (b) diffusion model, and (c) graph size}
    \label{fig:im}
    \centering
    \subfigure{\includegraphics[width=0.48\textwidth]{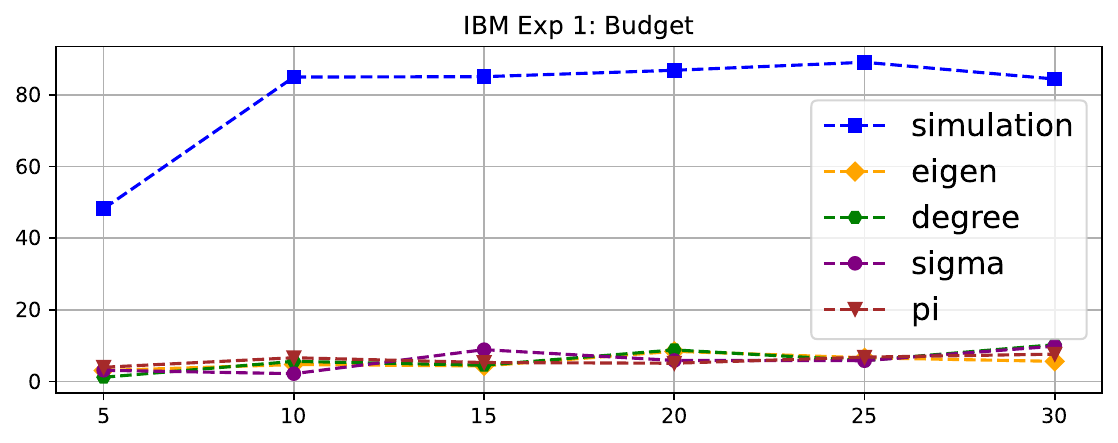}}
    \subfigure{\includegraphics[width=0.48\textwidth]{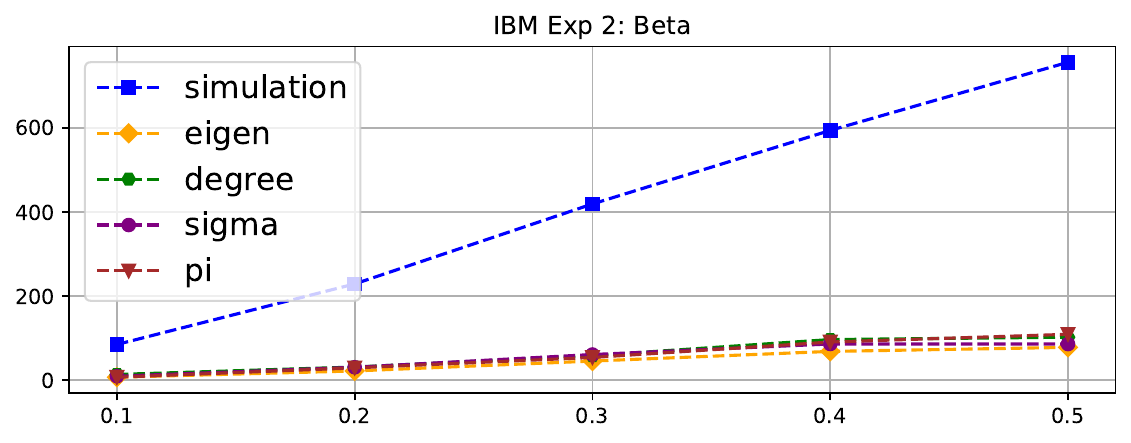}}
    \subfigure{\includegraphics[width=0.48\textwidth]{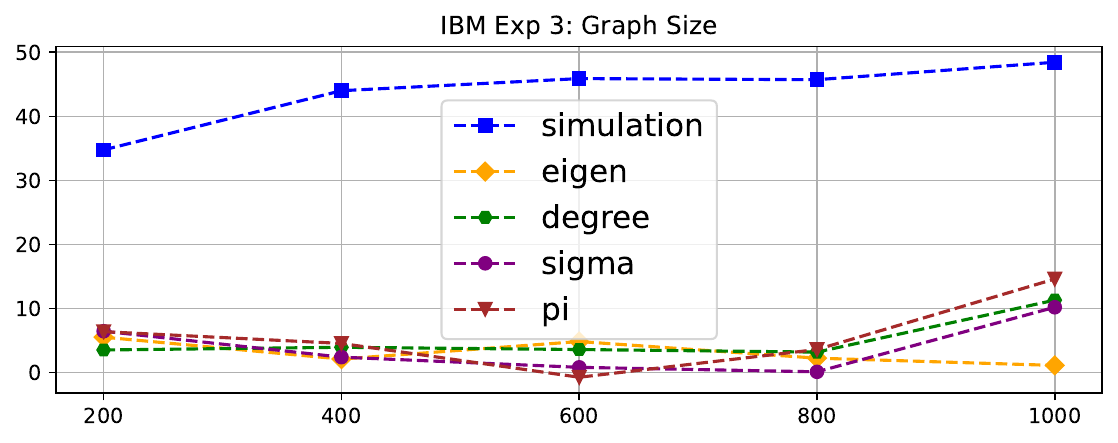}}
    \caption{The blocking effect regarding (a) budget (b) diffusion model, and (c) graph size}
    \label{fig:ibm}
\end{wrapfigure}
The baselines we tested in the experiments include CELF++, pi, degree, eigen-centrality, and RIS for IM and simulation-based greedy, pi, sigma, degree, and eigen-centrality for IBM. Metrics, including run-time, mean, and standard deviation, were recorded for further analysis. Figure~\ref{fig:im} and Figure~\ref{fig:ibm} show the influence spread regarding the three experiment settings for IM and IBM, respectively.

\textbf{SL.}
Our experimental setup encompassed a multi-source localization scenario, where we compared three state-of-the-art baseline algorithms: NETSLEUTH (implemented via two distinct methodologies), Jordan Centrality~\cite{7913632}, and LISN~\cite{osti_10125022}. We conducted tests under two different configurations. In the setup with two identified sources, we performed ten trials, and for the configuration involving three sources, we executed five trials. We employed three distinct graphs for these tests to further diversify our experimental conditions: CiteSeer, Cora, and WS-small world. Our data collection focused on the mean and standard deviation of the distances and the run time. These comprehensive results will facilitate detailed analysis and comparison of the performance of the tested algorithms.
\begin{wraptable}[9]{r}[0pt]{1.35\linewidth}
\fbox{
\scalebox{0.72}{
\begin{tabular}{l|r|r|r|r}
                                                                                              & \multicolumn{1}{c|}{Config} & \multicolumn{1}{c|}{CiteSeer} & \multicolumn{1}{c|}{Cora} & \multicolumn{1}{c}{WS-small world} \\ \hline
\begin{tabular}[c]{@{}l@{}}NETSLEUTH (legacy)\end{tabular}                               & 2 sources                   & 20.10$\pm$2.13                & 14.50$\pm$3.95            & 11.20$\pm$2.53                            \\ \hline
\multirow{2}{*}{\begin{tabular}[c]{@{}l@{}}NETSLEUTH\end{tabular}}        & 2 sources                   & 10.30$\pm$2.45                & 11.10$\pm$2.56            & 11.10$\pm$1.29                            \\ \cline{2-5} 
                                                                                              & 3 sources                   & 5.00$\pm$1.41                 & 9.20$\pm$0.84             & 7.60$\pm$1.52                             \\ \hline
\multirow{2}{*}{\begin{tabular}[c]{@{}l@{}}Jordan Centrality\end{tabular}} & 2 sources                   & 11.20$\pm$2.82                & 9.90$\pm$2.64             & 10.80$\pm$1.62                            \\ \cline{2-5} 
                                                                                              & 3 sources                   & 6.60$\pm$5.18                 & 7.60$\pm$2.70             & 8.00$\pm$0.00                             \\ \hline
\multirow{2}{*}{\begin{tabular}[c]{@{}l@{}}LISN\end{tabular}}              & 2 sources                   & 9.50$\pm$1.84                 & 10.20$\pm$4.34            & 10.70$\pm$1.16                            \\ \cline{2-5} 
                                                                                              & 3 sources                   & 3.60$\pm$2.51                 & 8.00$\pm$2.83             & 8.00$\pm$1.00                            
\end{tabular}%
}
}
\caption{Source localization performance measured by the distance between predicted and the actual sources.}
\label{tab:SL}
\end{wraptable}

\section{Maintenance and Development Plan}

\textbf{Next-phase development. }
One leading developer will response to general and publicly reported issues, while our team will provide additional human resources in the event of an escalated workload. Besides regular maintenance for the implemented components, in the forthcoming phase, our team will undertake several key tasks below. Firstly, we will prioritize the identification and analysis of crucial \textbf{tasks} within the field of network science that hold specific relevance to graph flow. These tasks encompass a diverse array of subjects, including but not limited to multilayer networks, higher-order networks, dynamic information maximization (IM), competitive IM, and more. See Section \ref{sec:future} for more details.
Secondly, our team will emphasize acquiring and integrating additional \textbf{real-world datasets}. Our ongoing projects are dedicated to datasets such as the genome-protein network, power network, as well as the structural and functional networks of the brain. These datasets serve as invaluable resources for advancing our research endeavors.
Thirdly, we plan to delve deeper into exploring diverse \textbf{diffusion models}. We aim to incorporate personalized propagation techniques to overcome the limitations associated with existing diffusion methods, which often struggle to accurately capture real-world interactions or achieve global optimality. The integration of personalized propagation may involve leveraging techniques from reinforcement learning and optimal transport methods, such as the network simplex algorithm.
Lastly, we will explore the implementation of \textbf{explanation methods} to provide a comprehensive range of insights into the outcomes of our methods. This will enable in-depth inspection and understanding of the underlying mechanisms and results.

\textbf{Dissemination.} In addition to conventional means of dissemination, such as publishing and presenting at conferences, our team intends to furnish relevant study materials and make them available on the \url{https://xflow.network} platform. The course material will encompass our latest research findings, relevant literature, and cutting-edge themes pertaining to graph flows. Our intention is to produce a scholarly survey paper utilizing this benchmark as a foundation. In addition, we intend to extend invitations to suitable collaborators who can provide pertinent datasets or baseline techniques.

\section{Discussion}\label{sec:future}

The present study's proposed framework has prompted the identification of several emerging directions that are outlined below.

\textit{\textbf{Opportunity 1: Flows over Multilayer Networks. }} 
Multilayer dynamics \cite{boccaletti2014structure,kivela2014multilayer,mucha2010spectral,glattfelder2011network,de2013centrality,wang2019coevolution,bianconi2018multilayer} studies complex systems with numerous interconnected layers or networks, each representing a particular relationship or interaction. Entities (nodes) can be connected over numerous levels, creating rich and diverse connectedness and dynamics.
Multilayer dynamics studies how system layers interact to produce emergent phenomena and complicated behaviors. It lets researchers examine multidimensional systems like social networks, transportation networks, biological networks, and infrastructure systems.
Multilayer dynamics entails understanding the interactions across layers, defining the dynamics and evolution of the system over time, and discovering the collective behaviors that result from these interactions. 

\textit{\textbf{Opportunity 2: Higher-order Flows.}} 
Higher-order flows analyze network interactions between groups of nodes \cite{battiston2020networks,battiston2021physics,li2022competing,boccaletti2023structure,ghorbanchian2021higher,fan2022epidemics,bianconi2021higher,parastesh2022synchronization,bick2021higher,zhang2023higher,alvarez2021evolutionary,bodnar2021weisfeiler,majhi2022dynamics,landry2023xgi}. Higher-order flow analysis captures the collective activity of several nodes, unlike pairwise network analysis.
Higher-order flows reveal patterns and dynamics that individual node interactions cannot. This technique helps understand complex systems including social, biological, and transportation networks, where collective behavior and group interactions are important.
It improves complex system modeling, prediction, and decision-making by helping comprehend how groups of nodes interact.

\textit{\textbf{Opportunity 3: Multi-Flows.}} 
Multi-flows over networks explore complicated graph flows and interactions \cite{wang2010networked,zhang2021emotional,wang2015coupled,pan2020phase,zino2020two,zhang2022interaction,peng2021multilayer,li2023coevolution,pinotti2020interplay,maghool2019coevolution}. Multi-flow analysis addresses the presence and interaction of various flow processes in a network, unlike traditional network analysis, which concentrates on a single flow (e.g., information flow, traffic flow).
Multi-flows understand that networks frequently entail several interactions, such as information sharing, resource allocation, and influence propagation, which can occur simultaneously and interact. Researchers examine multi-flows to understand the collective dynamics and complicated behavior resulting from multiple flow processes.

\newpage
\section*{Acknowledgement}
This work was funded by the NSF IIS award \# 2153369 and USDA-ARS 58-0200-0-002.
\bibliographystyle{unsrt}
\bibliography{z}

\begin{thebibliography}{100}

\bibitem{knoke2019social}
David Knoke and Song Yang.
\newblock {\em Social network analysis}.
\newblock SAGE publications, 2019.

\bibitem{li2021modeling}
Jian Li, Tao Xiang, and Linghui He.
\newblock Modeling epidemic spread in transportation networks: A review.
\newblock {\em Journal of Traffic and Transportation Engineering (English
  Edition)}, 8(2):139--152, 2021.

\bibitem{gao2020analysis}
Wenyou Gao and Chang Su.
\newblock Analysis on block chain financial transaction under artificial neural
  network of deep learning.
\newblock {\em Journal of Computational and Applied Mathematics}, 380:112991,
  2020.

\bibitem{boyaci2021graph}
Osman Boyaci, Amarachi Umunnakwe, Abhijeet Sahu, Mohammad~Rasoul Narimani,
  Muhammad Ismail, Katherine~R Davis, and Erchin Serpedin.
\newblock Graph neural networks based detection of stealth false data injection
  attacks in smart grids.
\newblock {\em IEEE Systems Journal}, 2021.

\bibitem{lynn2019physics}
Christopher~W Lynn and Danielle~S Bassett.
\newblock The physics of brain network structure, function and control.
\newblock {\em Nature Reviews Physics}, 1(5):318--332, 2019.

\bibitem{wu2020comprehensive}
Zonghan Wu, Shirui Pan, Fengwen Chen, Guodong Long, Chengqi Zhang, and S~Yu
  Philip.
\newblock A comprehensive survey on graph neural networks.
\newblock {\em IEEE transactions on neural networks and learning systems},
  32(1):4--24, 2020.

\bibitem{zhou2020graph}
Jie Zhou, Ganqu Cui, Shengding Hu, Zhengyan Zhang, Cheng Yang, Zhiyuan Liu,
  Lifeng Wang, Changcheng Li, and Maosong Sun.
\newblock Graph neural networks: A review of methods and applications.
\newblock {\em AI open}, 1:57--81, 2020.

\bibitem{guo2022systematic}
Xiaojie Guo and Liang Zhao.
\newblock A systematic survey on deep generative models for graph generation.
\newblock {\em IEEE Transactions on Pattern Analysis and Machine Intelligence},
  2022.

\bibitem{ahuja_network_1993}
Ravindra~K. Ahuja, Thomas~L. Magnanti, and James~B. Orlin.
\newblock {\em Network flows: theory, algorithms, and applications}.
\newblock Prentice Hall, Englewood Cliffs, N.J, 1993.

\bibitem{cormen_introduction_2009}
Thomas~H. Cormen, editor.
\newblock {\em Introduction to algorithms}.
\newblock MIT Press, Cambridge, Mass, 3rd ed edition, 2009.
\newblock OCLC: ocn311310321.

\bibitem{edmonds_theoretical_1972}
Jack Edmonds and Richard~M. Karp.
\newblock Theoretical {Improvements} in {Algorithmic} {Efficiency} for
  {Network} {Flow} {Problems}.
\newblock {\em Journal of the ACM}, 19(2):248--264, April 1972.

\bibitem{ford_maximal_1956}
Lester~Randolph Ford and Delbert~R Fulkerson.
\newblock Maximal flow through a network.
\newblock {\em Canadian journal of Mathematics}, 8:399--404, 1956.

\bibitem{domingos_mining_2001}
Pedro Domingos and Matt Richardson.
\newblock Mining the network value of customers.
\newblock In {\em Proceedings of the seventh {ACM} {SIGKDD} international
  conference on {Knowledge} discovery and data mining}, pages 57--66, San
  Francisco California, August 2001. ACM.

\bibitem{kempe_maximizing_2003}
David Kempe, Jon Kleinberg, and Éva Tardos.
\newblock Maximizing the spread of influence through a social network.
\newblock In {\em Proceedings of the ninth {ACM} {SIGKDD} international
  conference on {Knowledge} discovery and data mining}, pages 137--146,
  Washington, D.C., August 2003. ACM.

\bibitem{cohen_efficient_2003}
Reuven Cohen, Shlomo Havlin, and Daniel ben Avraham.
\newblock Efficient {Immunization} {Strategies} for {Computer} {Networks} and
  {Populations}.
\newblock {\em Physical Review Letters}, 91(24):247901, December 2003.

\bibitem{leskovec_cost-effective_2007}
Jure Leskovec, Andreas Krause, Carlos Guestrin, Christos Faloutsos, Jeanne
  VanBriesen, and Natalie Glance.
\newblock Cost-effective outbreak detection in networks.
\newblock In {\em Proceedings of the 13th {ACM} {SIGKDD} international
  conference on {Knowledge} discovery and data mining}, pages 420--429, San
  Jose California USA, August 2007. ACM.

\bibitem{Boccaletti2023May}
S.~Boccaletti, P.~De~Lellis, C.~I. del Genio, K.~Alfaro-Bittner, R.~Criado,
  S.~Jalan, and M.~Romance.
\newblock {The structure and dynamics of networks with higher order
  interactions}.
\newblock {\em Phys. Rep.}, 1018:1--64, May 2023.

\bibitem{boccaletti2014structure}
Stefano Boccaletti, Ginestra Bianconi, Regino Criado, Charo~I Del~Genio,
  Jes{\'u}s G{\'o}mez-Gardenes, Miguel Romance, Irene Sendina-Nadal, Zhen Wang,
  and Massimiliano Zanin.
\newblock The structure and dynamics of multilayer networks.
\newblock {\em Physics reports}, 544(1):1--122, 2014.

\bibitem{hu2020open}
Weihua Hu, Matthias Fey, Marinka Zitnik, Yuxiao Dong, Hongyu Ren, Bowen Liu,
  Michele Catasta, and Jure Leskovec.
\newblock Open graph benchmark: Datasets for machine learning on graphs.
\newblock {\em Advances in neural information processing systems},
  33:22118--22133, 2020.

\bibitem{nr-aaai15}
Ryan~A. Rossi and Nesreen~K. Ahmed.
\newblock The network data repository with interactive graph analytics and
  visualization.
\newblock In {\em Proceedings of the Twenty-Ninth AAAI Conference on Artificial
  Intelligence}, 2015.

\bibitem{snapnets}
Jure Leskovec and Andrej Krevl.
\newblock {SNAP Datasets}: {Stanford} large network dataset collection.
\newblock \url{http://snap.stanford.edu/data}, June 2014.

\bibitem{Fey/Lenssen/2019}
Matthias Fey and Jan~E. Lenssen.
\newblock Fast graph representation learning with {PyTorch Geometric}.
\newblock In {\em ICLR Workshop on Representation Learning on Graphs and
  Manifolds}, 2019.

\bibitem{wang2019deep}
Minjie Wang, Da~Zheng, Zihao Ye, Quan Gan, Mufei Li, Xiang Song, Jinjing Zhou,
  Chao Ma, Lingfan Yu, Yu~Gai, et~al.
\newblock Deep graph library: A graph-centric, highly-performant package for
  graph neural networks.
\newblock {\em arXiv preprint arXiv:1909.01315}, 2019.

\bibitem{JMLR:v22:21-0343}
Meng Liu, Youzhi Luo, Limei Wang, Yaochen Xie, Hao Yuan, Shurui Gui, Haiyang
  Yu, Zhao Xu, Jingtun Zhang, Yi~Liu, Keqiang Yan, Haoran Liu, Cong Fu, Bora~M
  Oztekin, Xuan Zhang, and Shuiwang Ji.
\newblock Dig: A turnkey library for diving into graph deep learning research.
\newblock {\em Journal of Machine Learning Research}, 22(240):1--9, 2021.

\bibitem{rozemberczki2021pytorch}
Benedek Rozemberczki, Paul Scherer, Yixuan He, George Panagopoulos, Alexander
  Riedel, Maria Astefanoaei, Oliver Kiss, Ferenc Beres, Guzman Lopez, Nicolas
  Collignon, and Rik Sarkar.
\newblock {PyTorch Geometric Temporal: Spatiotemporal Signal Processing with
  Neural Machine Learning Models}.
\newblock In {\em Proceedings of the 30th ACM International Conference on
  Information and Knowledge Management}, page 4564–4573, 2021.

\bibitem{tp3d}
Thomas Chaton, Chaulet Nicolas, Sofiane Horache, and Loic Landrieu.
\newblock Torch-points3d: A modular multi-task frameworkfor reproducible deep
  learning on 3d point clouds.
\newblock In {\em 2020 International Conference on 3D Vision (3DV)}. IEEE,
  2020.

\bibitem{8547527}
Siddharth Samsi, Vijay Gadepally, Michael Hurley, Michael Jones, Edward Kao,
  Sanjeev Mohindra, Paul Monticciolo, Albert Reuther, Steven Smith, William
  Song, Diane Staheli, and Jeremy Kepner.
\newblock Graphchallenge.org: Raising the bar on graph analytic performance.
\newblock In {\em 2018 IEEE High Performance extreme Computing Conference
  (HPEC)}, pages 1--7, 2018.

\bibitem{landry2023xgi}
Nicholas~W Landry, Maxime Lucas, Iacopo Iacopini, Giovanni Petri, Alice
  Schwarze, Alice Patania, and Leo Torres.
\newblock Xgi: A python package for higher-order interaction networks.
\newblock {\em Journal of Open Source Software}, 8(85):5162, 2023.

\bibitem{ndlib}
Alessandro Daniele, Giulio Cimini, Matteo Chinazzi, Albert-Laszlo Barabasi, and
  Nicola Perra.
\newblock The dynamic social network simulator.
\newblock {\em EPJ Data Science}, 5(1):1--28, 2016.

\bibitem{McCabe2022joss}
Lucas~H. McCabe.
\newblock cosasi: Graph diffusion source inference in python.
\newblock {\em Journal of Open Source Software}, 7(80):4894, 2022.

\bibitem{jenness2018epimodel}
Samuel~M Jenness, Steven~M Goodreau, and Martina Morris.
\newblock Epimodel: an r package for mathematical modeling of infectious
  disease over networks.
\newblock {\em Journal of statistical software}, 84, 2018.

\bibitem{leskovec2007cost}
Jure Leskovec, Andreas Krause, Carlos Guestrin, Christos Faloutsos, Jeanne
  VanBriesen, and Natalie Glance.
\newblock Cost-effective outbreak detection in networks.
\newblock In {\em Proceedings of the 13th ACM SIGKDD international conference
  on Knowledge discovery and data mining}, pages 420--429, 2007.

\bibitem{goyal2011celf++}
Amit Goyal, Wei Lu, and Laks~VS Lakshmanan.
\newblock Celf++ optimizing the greedy algorithm for influence maximization in
  social networks.
\newblock In {\em Proceedings of the 20th international conference companion on
  World wide web}, pages 47--48, 2011.

\bibitem{chen2013information}
Wei Chen, Laks~VS Lakshmanan, and Carlos Castillo.
\newblock Information and influence propagation in social networks.
\newblock {\em Synthesis Lectures on Data Management}, 5(4):1--177, 2013.

\bibitem{arora2017debunking}
Akhil Arora, Sainyam Galhotra, and Sayan Ranu.
\newblock Debunking the myths of influence maximization: An in-depth
  benchmarking study.
\newblock In {\em Proceedings of the 2017 ACM international conference on
  management of data}, pages 651--666, 2017.

\bibitem{tang2014influence}
Youze Tang, Xiaokui Xiao, and Yanchen Shi.
\newblock Influence maximization: Near-optimal time complexity meets practical
  efficiency.
\newblock In {\em Proceedings of the 2014 ACM SIGMOD international conference
  on Management of data}, pages 75--86, 2014.

\bibitem{page1999pagerank}
Lawrence Page, Sergey Brin, Rajeev Motwani, and Terry Winograd.
\newblock The pagerank citation ranking: Bringing order to the web.
\newblock Technical report, Stanford InfoLab, 1999.

\bibitem{zhong2018identifying}
Lin-Feng Zhong, Ming-Sheng Shang, Xiao-Long Chen, and Shi-Ming Cai.
\newblock Identifying the influential nodes via eigen-centrality from the
  differences and similarities of structure.
\newblock {\em Physica A: Statistical Mechanics and its Applications},
  510:77--82, 2018.

\bibitem{chen2009efficient}
Wei Chen, Yajun Wang, and Siyu Yang.
\newblock Efficient influence maximization in social networks.
\newblock In {\em Proceedings of the 15th ACM SIGKDD international conference
  on Knowledge discovery and data mining}, pages 199--208, 2009.

\bibitem{kimura2009blocking}
Masahiro Kimura, Kazumi Saito, and Hiroshi Motoda.
\newblock Blocking links to minimize contamination spread in a social network.
\newblock {\em ACM Transactions on Knowledge Discovery from Data (TKDD)},
  3(2):1--23, 2009.

\bibitem{chen2010scalable}
Wei Chen, Chi Wang, and Yajun Wang.
\newblock Scalable influence maximization for prevalent viral marketing in
  large-scale social networks.
\newblock In {\em Proceedings of the 16th ACM SIGKDD international conference
  on Knowledge discovery and data mining}, pages 1029--1038, 2010.

\bibitem{chen2010scalableLT}
Wei Chen, Yifei Yuan, and Li~Zhang.
\newblock Scalable influence maximization in social networks under the linear
  threshold model.
\newblock In {\em 2010 IEEE international conference on data mining}, pages
  88--97. IEEE, 2010.

\bibitem{tong2010vulnerability}
Hanghang Tong, B~Aditya Prakash, Charalampos Tsourakakis, Tina Eliassi-Rad,
  Christos Faloutsos, and Duen~Horng Chau.
\newblock On the vulnerability of large graphs.
\newblock In {\em 2010 IEEE International Conference on Data Mining}, pages
  1091--1096. IEEE, 2010.

\bibitem{goyal2011simpath}
Amit Goyal, Wei Lu, and Laks~VS Lakshmanan.
\newblock Simpath: An efficient algorithm for influence maximization under the
  linear threshold model.
\newblock In {\em 2011 IEEE 11th international conference on data mining},
  pages 211--220. IEEE, 2011.

\bibitem{yan2019minimizing}
Ruidong Yan, Deying Li, Weili Wu, Ding-Zhu Du, and Yongcai Wang.
\newblock Minimizing influence of rumors by blockers on social networks:
  algorithms and analysis.
\newblock {\em IEEE Transactions on Network Science and Engineering},
  7(3):1067--1078, 2019.

\bibitem{zhang2022blocking}
Zonghan Zhang, Subhodip Biswas, Fanglan Chen, Kaiqun Fu, Taoran Ji, Chang-Tien
  Lu, Naren Ramakrishnan, and Zhiqian Chen.
\newblock Blocking influence at collective level with hard constraints (student
  abstract).
\newblock 2022.

\bibitem{budak2011limiting}
Ceren Budak, Divyakant Agrawal, and Amr El~Abbadi.
\newblock Limiting the spread of misinformation in social networks.
\newblock In {\em Proceedings of the 20th international conference on World
  wide web}, pages 665--674, 2011.

\bibitem{he2012influence}
Xinran He, Guojie Song, Wei Chen, and Qingye Jiang.
\newblock Influence blocking maximization in social networks under the
  competitive linear threshold model.
\newblock In {\em Proceedings of the 2012 siam international conference on data
  mining}, pages 463--474. SIAM, 2012.

\bibitem{ibrahim2018controlling}
Ragia~A Ibrahim, Hesham~A Hefny, and Aboul~Ella Hassanien.
\newblock Controlling social information cascade: a survey.
\newblock {\em Big Data Analytics}, pages 196--212, 2018.

\bibitem{fan2013least}
Lidan Fan, Zaixin Lu, Weili Wu, Bhavani Thuraisingham, Huan Ma, and Yuanjun Bi.
\newblock Least cost rumor blocking in social networks.
\newblock In {\em 2013 IEEE 33rd International Conference on Distributed
  Computing Systems}, pages 540--549. IEEE, 2013.

\bibitem{allcott2017social}
Hunt Allcott and Matthew Gentzkow.
\newblock Social media and fake news in the 2016 election.
\newblock {\em Journal of economic perspectives}, 31(2):211--36, 2017.

\bibitem{chowdhury2020joint}
Rajdipa Chowdhury, Sriram Srinivasan, and Lise Getoor.
\newblock Joint estimation of user and publisher credibility for fake news
  detection.
\newblock In {\em Proceedings of the 29th ACM International Conference on
  Information \& Knowledge Management}, pages 1993--1996, 2020.

\bibitem{ozili2020spillover}
Peterson~K Ozili and Thankom Arun.
\newblock Spillover of covid-19: impact on the global economy.
\newblock {\em Available at SSRN 3562570}, 2020.

\bibitem{sher2020impact}
Leo Sher.
\newblock The impact of the covid-19 pandemic on suicide rates.
\newblock {\em QJM: An International Journal of Medicine}, 113(10):707--712,
  2020.

\bibitem{area2017ebola}
Ivan Area, Faical Ndairou, Juan~J Nieto, Cristiana~J Silva, and Delfim~FM
  Torres.
\newblock Ebola model and optimal control with vaccination constraints.
\newblock {\em arXiv preprint arXiv:1703.01368}, 2017.

\bibitem{wang2013negative}
Senzhang Wang, Xiaojian Zhao, Yan Chen, Zhoujun Li, Kai Zhang, and Jiali Xia.
\newblock Negative influence minimizing by blocking nodes in social networks.
\newblock In {\em Proceedings of the 17th AAAI Conference on Late-Breaking
  Developments in the Field of Artificial Intelligence}, pages 134--136, 2013.

\bibitem{shi2019adaptive}
Qihao Shi, Can Wang, Deshi Ye, Jiawei Chen, Yan Feng, and Chun Chen.
\newblock Adaptive influence blocking: Minimizing the negative spread by
  observation-based policies.
\newblock In {\em 2019 IEEE 35th International Conference on Data Engineering
  (ICDE)}, pages 1502--1513. IEEE, 2019.

\bibitem{medlock2009optimizing}
Jan Medlock and Alison~P Galvani.
\newblock Optimizing influenza vaccine distribution.
\newblock {\em Science}, 325(5948):1705--1708, 2009.

\bibitem{clem2011fundamentals}
Angela~S Clem.
\newblock Fundamentals of vaccine immunology.
\newblock {\em Journal of global infectious diseases}, 3(1):73, 2011.

\bibitem{glass2006targeted}
Robert~J Glass, Laura~M Glass, Walter~E Beyeler, and H~Jason Min.
\newblock Targeted social distancing designs for pandemic influenza.
\newblock {\em Emerging infectious diseases}, 12(11):1671, 2006.

\bibitem{kaur2017blocking}
Harneet Kaur and Jing He.
\newblock Blocking negative influential node set in social networks: from host
  perspective.
\newblock {\em Transactions on Emerging Telecommunications Technologies},
  28(4):e3007, 2017.

\bibitem{ma2016identifying}
Ling-ling Ma, Chuang Ma, Hai-Feng Zhang, and Bing-Hong Wang.
\newblock Identifying influential spreaders in complex networks based on
  gravity formula.
\newblock {\em Physica A: Statistical Mechanics and its Applications},
  451:205--212, 2016.

\bibitem{tong2012gelling}
Hanghang Tong, B~Aditya Prakash, Tina Eliassi-Rad, Michalis Faloutsos, and
  Christos Faloutsos.
\newblock Gelling, and melting, large graphs by edge manipulation.
\newblock In {\em Proceedings of the 21st ACM international conference on
  Information and knowledge management}, pages 245--254, 2012.

\bibitem{tong2017efficient}
Guangmo Tong, Weili Wu, Ling Guo, Deying Li, Cong Liu, Bin Liu, and Ding-Zhu
  Du.
\newblock An efficient randomized algorithm for rumor blocking in online social
  networks.
\newblock {\em IEEE Transactions on Network Science and Engineering},
  7(2):845--854, 2017.

\bibitem{jiang2016identifying}
Jiaojiao Jiang, Sheng Wen, Shui Yu, Yang Xiang, and Wanlei Zhou.
\newblock Identifying propagation sources in networks: State-of-the-art and
  comparative studies.
\newblock {\em IEEE Communications Surveys \& Tutorials}, 19(1):465--481, 2016.

\bibitem{shelke2019source}
Sushila Shelke and Vahida Attar.
\newblock Source detection of rumor in social network--a review.
\newblock {\em Online Social Networks and Media}, 9:30--42, 2019.

\bibitem{dong2019multiple}
Ming Dong, Bolong Zheng, Nguyen Quoc Viet~Hung, Han Su, and Guohui Li.
\newblock Multiple rumor source detection with graph convolutional networks.
\newblock In {\em Proceedings of the 28th ACM international conference on
  information and knowledge management}, pages 569--578, 2019.

\bibitem{prakash2012spotting}
B~Aditya Prakash, Jilles Vreeken, and Christos Faloutsos.
\newblock Spotting culprits in epidemics: How many and which ones?
\newblock In {\em 2012 IEEE 12th International Conference on Data Mining},
  pages 11--20. IEEE, 2012.

\bibitem{wang2017multiple}
Zheng Wang, Chaokun Wang, Jisheng Pei, and Xiaojun Ye.
\newblock Multiple source detection without knowing the underlying propagation
  model.
\newblock In {\em Proceedings of the AAAI Conference on Artificial
  Intelligence}, volume~31, 2017.

\bibitem{zhu2017catch}
Kai Zhu, Zhen Chen, and Lei Ying.
\newblock Catch’em all: Locating multiple diffusion sources in networks with
  partial observations.
\newblock In {\em Proceedings of the AAAI Conference on Artificial
  Intelligence}, volume~31, 2017.

\bibitem{zhu2014information}
Kai Zhu and Lei Ying.
\newblock Information source detection in the sir model: A sample-path-based
  approach.
\newblock {\em IEEE/ACM Transactions on Networking}, 24(1):408--421, 2014.

\bibitem{zang2015locating}
Wenyu Zang, Peng Zhang, Chuan Zhou, and Li~Guo.
\newblock Locating multiple sources in social networks under the sir model: A
  divide-and-conquer approach.
\newblock {\em Journal of Computational Science}, 10:278--287, 2015.

\bibitem{jiang2022graph}
Weiwei Jiang and Jiayun Luo.
\newblock Graph neural network for traffic forecasting: A survey.
\newblock {\em Expert Systems with Applications}, page 117921, 2022.

\bibitem{jin2023spatio}
Guangyin Jin, Yuxuan Liang, Yuchen Fang, Jincai Huang, Junbo Zhang, and
  Yu~Zheng.
\newblock Spatio-temporal graph neural networks for predictive learning in
  urban computing: A survey.
\newblock {\em arXiv preprint arXiv:2303.14483}, 2023.

\bibitem{goldenberg_talk_2001}
Jacob Goldenberg, Barak Libai, and Eitan Muller.
\newblock Talk of the {Network}: {A} {Complex} {Systems} {Look} at the
  {Underlying} {Process} of {Word}-of-{Mouth}.
\newblock {\em Marketing Letters}, 12(3):211--223, August 2001.

\bibitem{granovetter_threshold_1978}
Mark Granovetter.
\newblock Threshold {Models} of {Collective} {Behavior}.
\newblock {\em American Journal of Sociology}, 83(6):1420--1443, May 1978.

\bibitem{kermack_contribution_1927}
W.~O. Kermack and A.~G. McKendrick.
\newblock A {Contribution} to the {Mathematical} {Theory} of {Epidemics}.
\newblock {\em Proceedings of the Royal Society of London. Series A, Containing
  Papers of a Mathematical and Physical Character}, 115(772):700--721, 1927.

\bibitem{Kermack1927}
W.~O. Kermack and A.~G. McKendrick.
\newblock A contribution to the mathematical theory of epidemics.
\newblock {\em Proceedings of the Royal Society of London. Series A, Containing
  Papers of a Mathematical and Physical Character}, 115(772):700--721, 1927.

\bibitem{Anderson1980}
Roy~M. Anderson and Robert~M. May.
\newblock The population dynamics of microparasites and their invertebrate
  hosts.
\newblock {\em Philosophical Transactions of the Royal Society of London.
  Series B, Biological Sciences}, 291(1054):451--524, 1980.

\bibitem{Hethcote2000}
Herbert~W. Hethcote.
\newblock The mathematics of infectious diseases.
\newblock {\em SIAM Review}, 42(4):599--653, 2000.

\bibitem{Heesterbeek1996}
Hans Heesterbeek, Roy~M. Anderson, Viggo Andreasen, Shweta Bansal, Daniela
  De~Angelis, Christopher Dye, Ken Eames, W.~John Edmunds, Simon Frost,
  Sebastian Funk, et~al.
\newblock Concepts and methodological issues in infectious disease modeling.
\newblock {\em Epidemiology and Infection}, 133(02):389--403, 2005.

\bibitem{guo2013personalized}
Jing Guo, Peng Zhang, Chuan Zhou, Yanan Cao, and Li~Guo.
\newblock Personalized influence maximization on social networks.
\newblock In {\em Proceedings of the 22nd ACM international conference on
  Information \& Knowledge Management}, pages 199--208, 2013.

\bibitem{li2015real}
Yuchen Li, Dongxiang Zhang, and Kian-Lee Tan.
\newblock Real-time targeted influence maximization for online advertisements.
\newblock 2015.

\bibitem{nguyen2016cost}
Hung~T Nguyen, Thang~N Dinh, and My~T Thai.
\newblock Cost-aware targeted viral marketing in billion-scale networks.
\newblock In {\em IEEE INFOCOM 2016-the 35th annual IEEE international
  conference on computer communications}, pages 1--9. IEEE, 2016.

\bibitem{chen2016real}
Wei Chen, Tian Lin, and Cheng Yang.
\newblock Real-time topic-aware influence maximization using preprocessing.
\newblock {\em Computational social networks}, 3(1):1--19, 2016.

\bibitem{li2017discovering}
Yuchen Li, Ju~Fan, Dongxiang Zhang, and Kian-Lee Tan.
\newblock Discovering your selling points: Personalized social influential tags
  exploration.
\newblock In {\em Proceedings of the 2017 ACM International Conference on
  Management of Data}, pages 619--634, 2017.

\bibitem{tian2020deep}
Shan Tian, Songsong Mo, Liwei Wang, and Zhiyong Peng.
\newblock Deep reinforcement learning-based approach to tackle topic-aware
  influence maximization.
\newblock {\em Data Science and Engineering}, 5:1--11, 2020.

\bibitem{hagberg2008exploring}
Aric Hagberg, Pieter Swart, and Daniel S~Chult.
\newblock Exploring network structure, dynamics, and function using networkx.
\newblock Technical report, Los Alamos National Lab.(LANL), Los Alamos, NM
  (United States), 2008.

\bibitem{cohen2014sketch}
Edith Cohen, Daniel Delling, Thomas Pajor, and Renato~F Werneck.
\newblock Sketch-based influence maximization and computation: Scaling up with
  guarantees.
\newblock In {\em Proceedings of the 23rd ACM international conference on
  conference on information and knowledge management}, pages 629--638, 2014.

\bibitem{tang2015influence}
Youze Tang, Yanchen Shi, and Xiaokui Xiao.
\newblock Influence maximization in near-linear time: A martingale approach.
\newblock In {\em Proceedings of the 2015 ACM SIGMOD international conference
  on management of data}, pages 1539--1554, 2015.

\bibitem{borgs2014maximizing}
Christian Borgs, Michael Brautbar, Jennifer Chayes, and Brendan Lucier.
\newblock Maximizing social influence in nearly optimal time.
\newblock In {\em Proceedings of the twenty-fifth annual ACM-SIAM symposium on
  Discrete algorithms}, pages 946--957. SIAM, 2014.

\bibitem{6413787}
B.~Aditya Prakash, Jilles Vreeken, and Christos Faloutsos.
\newblock Spotting culprits in epidemics: How many and which ones?
\newblock In {\em 2012 IEEE 12th International Conference on Data Mining},
  pages 11--20, 2012.

\bibitem{7913632}
Wuqiong Luo, Wee~Peng Tay, and Mei Leng.
\newblock On the universality of jordan centers for estimating infection
  sources in tree networks.
\newblock {\em IEEE Transactions on Information Theory}, 63(7):4634--4657,
  2017.

\bibitem{osti_10125022}
Guanyu Nie and Christopher Quinn.
\newblock Localizing the information source in a network.
\newblock {\em TrueFact 2019 : KDD 2019 Workshop on Truth Discovery and Fact
  Checking: Theory and Practice}.

\bibitem{kivela2014multilayer}
Mikko Kivela, Alex Arenas, Marc Barthelemy, James~P. Gleeson, Yamir Moreno, and
  Mason~A. Porter.
\newblock Multilayer networks.
\newblock {\em Journal of Complex Networks}, 2014.

\bibitem{mucha2010spectral}
Peter~J. Mucha, Thomas Richardson, Kevin Macon, Mason~A. Porter, and
  Jukka-Pekka Onnela.
\newblock Spectral clustering of multilayer networks.
\newblock {\em Physical Review E}, 2010.

\bibitem{glattfelder2011network}
James~B. Glattfelder and Stefano Battiston.
\newblock The network of global corporate control.
\newblock {\em PLoS ONE}, 2011.

\bibitem{de2013centrality}
Manlio De~Domenico, Vincenzo Nicosia, Alex Arenas, and Vito Latora.
\newblock Centrality measures and degree correlations in multilayer networks.
\newblock {\em Physical Review Letters}, 2013.

\bibitem{wang2019coevolution}
Wei Wang, Quan-Hui Liu, Junhao Liang, Yanqing Hu, and Tao Zhou.
\newblock Coevolution spreading in complex networks.
\newblock {\em Physics Reports}, 820:1--51, 2019.

\bibitem{bianconi2018multilayer}
Ginestra Bianconi.
\newblock {\em Multilayer networks: structure and function}.
\newblock Oxford university press, 2018.

\bibitem{battiston2020networks}
Federico Battiston, Giulia Cencetti, Iacopo Iacopini, Vito Latora, Maxime
  Lucas, Alice Patania, Jean-Gabriel Young, and Giovanni Petri.
\newblock Networks beyond pairwise interactions: structure and dynamics.
\newblock {\em Physics Reports}, 874:1--92, 2020.

\bibitem{battiston2021physics}
Federico Battiston, Enrico Amico, Alain Barrat, Ginestra Bianconi, Guilherme
  Ferraz~de Arruda, Benedetta Franceschiello, Iacopo Iacopini, Sonia K{\'e}fi,
  Vito Latora, Yamir Moreno, et~al.
\newblock The physics of higher-order interactions in complex systems.
\newblock {\em Nature Physics}, 17(10):1093--1098, 2021.

\bibitem{li2022competing}
WenYao Li, Xiaoyu Xue, Liming Pan, Tao Lin, and Wei Wang.
\newblock Competing spreading dynamics in simplicial complex.
\newblock {\em Applied Mathematics and Computation}, 412:126595, 2022.

\bibitem{boccaletti2023structure}
Stefano Boccaletti, Pietro De~Lellis, CI~del Genio, Karin Alfaro-Bittner,
  Regino Criado, Sarika Jalan, and Miguel Romance.
\newblock The structure and dynamics of networks with higher order
  interactions.
\newblock {\em Physics Reports}, 1018:1--64, 2023.

\bibitem{ghorbanchian2021higher}
Reza Ghorbanchian, Juan~G Restrepo, Joaqu{\'\i}n~J Torres, and Ginestra
  Bianconi.
\newblock Higher-order simplicial synchronization of coupled topological
  signals.
\newblock {\em Communications Physics}, 4(1):120, 2021.

\bibitem{fan2022epidemics}
Junfeng Fan, Qian Yin, Chengyi Xia, and Matja{\v{z}} Perc.
\newblock Epidemics on multilayer simplicial complexes.
\newblock {\em Proceedings of the Royal Society A}, 478(2261):20220059, 2022.

\bibitem{bianconi2021higher}
Ginestra Bianconi.
\newblock {\em Higher-order networks}.
\newblock Cambridge University Press, 2021.

\bibitem{parastesh2022synchronization}
Fatemeh Parastesh, Mahtab Mehrabbeik, Karthikeyan Rajagopal, Sajad Jafari, and
  Matja{\v{z}} Perc.
\newblock Synchronization in hindmarsh--rose neurons subject to higher-order
  interactions.
\newblock {\em Chaos: An Interdisciplinary Journal of Nonlinear Science},
  32(1):013125, 2022.

\bibitem{bick2021higher}
Christian Bick, Elizabeth Gross, Heather~A Harrington, and Michael~T Schaub.
\newblock What are higher-order networks?
\newblock {\em arXiv preprint arXiv:2104.11329}, 2021.

\bibitem{zhang2023higher}
Yuanzhao Zhang, Maxime Lucas, and Federico Battiston.
\newblock Higher-order interactions shape collective dynamics differently in
  hypergraphs and simplicial complexes.
\newblock {\em Nature Communications}, 14(1):1605, 2023.

\bibitem{alvarez2021evolutionary}
Unai Alvarez-Rodriguez, Federico Battiston, Guilherme~Ferraz de~Arruda, Yamir
  Moreno, Matja{\v{z}} Perc, and Vito Latora.
\newblock Evolutionary dynamics of higher-order interactions in social
  networks.
\newblock {\em Nature Human Behaviour}, 5(5):586--595, 2021.

\bibitem{bodnar2021weisfeiler}
Cristian Bodnar, Fabrizio Frasca, Yuguang Wang, Nina Otter, Guido~F Montufar,
  Pietro Lio, and Michael Bronstein.
\newblock Weisfeiler and lehman go topological: Message passing simplicial
  networks.
\newblock In {\em International Conference on Machine Learning}, pages
  1026--1037. PMLR, 2021.

\bibitem{majhi2022dynamics}
Soumen Majhi, Matja{\v{z}} Perc, and Dibakar Ghosh.
\newblock Dynamics on higher-order networks: A review.
\newblock {\em Journal of the Royal Society Interface}, 19(188):20220043, 2022.

\bibitem{wang2010networked}
Yingchun Wang, Huaguang Zhang, Xingyuan Wang, and Dongsheng Yang.
\newblock Networked synchronization control of coupled dynamic networks with
  time-varying delay.
\newblock {\em IEEE Transactions on Systems, Man, and Cybernetics, Part B
  (Cybernetics)}, 40(6):1468--1479, 2010.

\bibitem{zhang2021emotional}
Guijuan Zhang, Dianjie Lu, and Xiaohua Jia.
\newblock Emotional contagion in physical--cyber integrated networks: The phase
  transition perspective.
\newblock {\em IEEE Transactions on Cybernetics}, 52(8):7875--7888, 2021.

\bibitem{wang2015coupled}
Zhen Wang, Michael~A Andrews, Zhi-Xi Wu, Lin Wang, and Chris~T Bauch.
\newblock Coupled disease--behavior dynamics on complex networks: A review.
\newblock {\em Physics of life reviews}, 15:1--29, 2015.

\bibitem{pan2020phase}
Liming Pan, Dan Yang, Wei Wang, Shimin Cai, Tao Zhou, and Ying-Cheng Lai.
\newblock Phase diagrams of interacting spreading dynamics in complex networks.
\newblock {\em Physical Review Research}, 2(2):023233, 2020.

\bibitem{zino2020two}
Lorenzo Zino, Mengbin Ye, and Ming Cao.
\newblock A two-layer model for coevolving opinion dynamics and collective
  decision-making in complex social systems.
\newblock {\em Chaos: An Interdisciplinary Journal of Nonlinear Science},
  30(8):083107, 2020.

\bibitem{zhang2022interaction}
Hegui Zhang, Xiaolong Chen, Yi~Peng, Gang Kou, and Ruijie Wang.
\newblock The interaction of multiple information on multiplex social networks.
\newblock {\em Information Sciences}, 605:366--380, 2022.

\bibitem{peng2021multilayer}
Kaiyan Peng, Zheng Lu, Vanessa Lin, Michael~R Lindstrom, Christian Parkinson,
  Chuntian Wang, Andrea~L Bertozzi, and Mason~A Porter.
\newblock A multilayer network model of the coevolution of the spread of a
  disease and competing opinions.
\newblock {\em Mathematical Models and Methods in Applied Sciences},
  31(12):2455--2494, 2021.

\bibitem{li2023coevolution}
Wenyao Li, Meng Cai, Xiaoni Zhong, Yanbing Liu, Tao Lin, and Wei Wang.
\newblock Coevolution of epidemic and infodemic on higher-order networks.
\newblock {\em Chaos, Solitons \& Fractals}, 168:113102, 2023.

\bibitem{pinotti2020interplay}
Francesco Pinotti, Fakhteh Ghanbarnejad, Philipp H{\"o}vel, and Chiara Poletto.
\newblock Interplay between competitive and cooperative interactions in a
  three-player pathogen system.
\newblock {\em Royal Society open science}, 7(1):190305, 2020.

\bibitem{maghool2019coevolution}
Samira Maghool, Nahid Maleki-Jirsaraei, and Marco Cremonini.
\newblock The coevolution of contagion and behavior with increasing and
  decreasing awareness.
\newblock {\em PloS one}, 14(12):e0225447, 2019.

\end{thebibliography}

\newpage
\section*{Checklist}


\begin{enumerate}

\item For all authors...
\begin{enumerate}
  \item Do the main claims made in the abstract and introduction accurately reflect the paper's contributions and scope?
    \answerYes{}
  \item Did you describe the limitations of your work?
    \answerYes{}
  \item Did you discuss any potential negative societal impacts of your work?
    \answerYes{}
  \item Have you read the ethics review guidelines and ensured that your paper conforms to them?
    \answerYes{}
\end{enumerate}

\item If you are including theoretical results...
\begin{enumerate}
  \item Did you state the full set of assumptions of all theoretical results?
    \answerNA{}
	\item Did you include complete proofs of all theoretical results?
    \answerNA{}
\end{enumerate}

\item If you ran experiments (e.g. for benchmarks)...
\begin{enumerate}
  \item Did you include the code, data, and instructions needed to reproduce the main experimental results (either in the supplemental material or as a URL)?
    \answerYes{}
  \item Did you specify all the training details (e.g., data splits, hyperparameters, how they were chosen)?
    \answerYes{All details are in the code and the website, description will not in the main submission due to the page limit.}
	\item Did you report error bars (e.g., with respect to the random seed after running experiments multiple times)?
    \answerYes{Mean is report in the main file, and variance will be provided in the supplementary material and website}
	\item Did you include the total amount of compute and the type of resources used (e.g., type of GPUs, internal cluster, or cloud provider)?
    \answerYes{}
\end{enumerate}

\item If you are using existing assets (e.g., code, data, models) or curating/releasing new assets...
\begin{enumerate}
  \item If your work uses existing assets, did you cite the creators?
    \answerYes{See related work section}
  \item Did you mention the license of the assets?
    \answerYes{All official references are provided, inside which license are listed}
  \item Did you include any new assets either in the supplemental material or as a URL?
    \answerNA{Our project is a benchmark suite without new datasets}
  \item Did you discuss whether and how consent was obtained from people whose data you're using/curating?
    \answerNA{all datasets are publicly available, we are building benchmarks beyond them}
  \item Did you discuss whether the data you are using/curating contains personally identifiable information or offensive content?
    \answerNA{no personally identifiable information or offensive content included}
\end{enumerate}

\item If you used crowdsourcing or conducted research with human subjects...
\begin{enumerate}
  \item Did you include the full text of instructions given to participants and screenshots, if applicable?
    \answerNA{no crowdsourcing or conducted research with human subjects}
  \item Did you describe any potential participant risks, with links to Institutional Review Board (IRB) approvals, if applicable?
    \answerNA{no crowdsourcing or conducted research with human subjects}
  \item Did you include the estimated hourly wage paid to participants and the total amount spent on participant compensation?
    \answerNA{no crowdsourcing or conducted research with human subjects}
\end{enumerate}

\end{enumerate}

\newpage
\appendix

\section{Appendix}

\subsection{Answers to Official Questions}
Include extra information in the appendix. This section will often be part of the supplemental material. Please see the call on the NeurIPS website for links to additional guides on dataset publication.

\begin{enumerate}

\item Submission introducing new datasets must include the following in the supplementary materials:
\begin{enumerate}
  \item Dataset documentation and intended uses. Recommended documentation frameworks include datasheets for datasets, dataset nutrition labels, data statements for NLP, and accountability frameworks. 
  \begin{itemize}
      \item \textit{no dataset will be introduced, but our documentation can be found in our  portal URL address at \url{https://github.com/XGraphing/XFlow}}
  \end{itemize}
  \item URL to website/platform where the dataset/benchmark can be viewed and downloaded by the reviewers.
  \begin{itemize}
      \item \textit{See \url{https://github.com/XGraphing/XFlow}}
  \end{itemize}
  \item Author statement that they bear all responsibility in case of violation of rights, etc., and confirmation of the data license.
  \begin{itemize}
      \item \textit{We bear all responsibility in case of violation of rights, see more details including license in \url{https://github.com/XGraphing/XFlow}}
  \end{itemize}
  \item Hosting, licensing, and maintenance plan. The choice of hosting platform is yours, as long as you ensure access to the data (possibly through a curated interface) and will provide the necessary maintenance.
  \begin{itemize}
      \item \textit{We will use GitHub and perform our maintenance}
  \end{itemize}
  
\end{enumerate}

\item To ensure accessibility, the supplementary materials for datasets must include the following:
\begin{enumerate}
  \item Links to access the dataset and its metadata. This can be hidden upon submission if the dataset is not yet publicly available but must be added in the camera-ready version. In select cases, e.g when the data can only be released at a later date, this can be added afterward. Simulation environments should link to (open source) code repositories.
  \begin{itemize}
      \item \textit{We don't provide new datasets, but the references of the datasets we used in the benchmark are provided}
  \end{itemize}
  \item The dataset itself should ideally use an open and widely used data format. Provide a detailed explanation on how the dataset can be read. For simulation environments, use existing frameworks or explain how they can be used.
  \begin{itemize}
      \item \textit{We don't provide new datasets, and we provide the dataset in the original format in dependency libraries. The dataset format are widely-accepted, such as \hyperlink{https://networkx.org/}{NetworkX} graph object and \hyperlink{https://pytorch-geometric.readthedocs.io}{Pytorch Geometric} graph object. We also have simulation envionment which mainly utilize existing tool \hyperlink{https://ndlib.readthedocs.io/en/latest/}{NDLib}. }
  \end{itemize}
  \item Long-term preservation: It must be clear that the dataset will be available for a long time, either by uploading to a data repository or by explaining how the authors themselves will ensure this.
  \begin{itemize}
      \item \textit{We don't provide new datasets, but use existing one, and the users can also use their owns.}
  \end{itemize}
  \item Explicit license: Authors must choose a license, ideally a CC license for datasets, or an open source license for code (e.g. RL environments).
    \begin{itemize}
      \item \textit{We will use MIT License.}
  \end{itemize}
  \item Add structured metadata to a dataset's meta-data page using Web standards (like schema.org and DCAT): This allows it to be discovered and organized by anyone. If you use an existing data repository, this is often done automatically.
  \begin{itemize}
      \item \textit{We don't provide new datasets, but use existing one, and the users can also use their owns.}
  \end{itemize}
  \item Highly recommended: a persistent dereferenceable identifier (e.g. a DOI minted by a data repository or a prefix on identifiers.org) for datasets, or a code repository (e.g. GitHub, GitLab,...) for code. If this is not possible or useful, please explain why.
    \begin{itemize}
      \item \textit{We use GitHub \url{https://github.com/XGraphing/XFlow}}
  \end{itemize}
\end{enumerate}

\item For benchmarks, the supplementary materials must ensure that all results are easily reproducible. Where possible, use a reproducibility framework such as the ML reproducibility checklist, or otherwise guarantee that all results can be easily reproduced, i.e. all necessary datasets, code, and evaluation procedures must be accessible and documented.
    \begin{itemize}
      \item \textit{We provide example code and necessary documentation.}
  \end{itemize}
\item For papers introducing best practices in creating or curating datasets and benchmarks, the above supplementary materials are not required.
\end{enumerate}

\subsection{Error Bar}
Table \ref{tab:1}-\ref{tab:3} lists the mean and variance of the influence expectation (IE) under various IM configurations. 
Table \ref{tab:4}-\ref{tab:6} lists the mean and variance of the blocking effect (blocked) under various IBM configurations. 
Table \ref{tab:7} lists the mean and variance of the top score distance under various SL configurations. 
The table data corresponds to Figure \ref{fig:im}, \ref{fig:ibm} and Table \ref{tab:SL},

\begin{table}[hpb]
\scalebox{0.8}{
\begin{tabular}{|c|c|c|c|c|c|c|c|}
\hline
\multirow{3}{*}{Baselines} & beta       & 0.1             & 0.1              & 0.1              & 0.1              & 0.1              & 0.1              \\ \cline{2-8} 
                           & budget     & 5               & 10               & 15               & 20               & 25               & 30               \\ \cline{2-8} 
                           & graph size & 1000            & 1000             & 1000             & 1000             & 1000             & 1000             \\ \hline
pi                         & IE         & 65.34$\pm$13.26 & 129.92$\pm$19.22 & 178.12$\pm$19.76 & 228.29$\pm$22.52 & 275.01$\pm$23.77 & 305.98$\pm$23.82 \\ \hline
degree                     & IE         & 66.89$\pm$13.12 & 129.05$\pm$19.22 & 177.91$\pm$21.36 & 228.76$\pm$22.2  & 274.31$\pm$23.12 & 307.23$\pm$23.52 \\ \hline
eigen                      & IE         & 63$\pm$13.04    & 129.24$\pm$19.22 & 181.61$\pm$20.85 & 231.82$\pm$21.59 & 281.25$\pm$24.37 & 332.42$\pm$26.43 \\ \hline
RIS                        & IE         & 65.05$\pm$12.92 & 120.04$\pm$17.44 & 167.83$\pm$19.73 & 236.96$\pm$24.7  & 272.57$\pm$23.81 & 316.75$\pm$24.54 \\ \hline
CELF++                     & IE         & 63.99$\pm$13.7  & 120.88$\pm$19.22 & 175.22$\pm$21.31 & 218.57$\pm$22.89 & 262.96$\pm$23.88 & 310.61$\pm$24.9  \\ \hline
\end{tabular}
}
\caption{Error Bar of IM Exp 1: Budget}
\label{tab:1}
\end{table}

\begin{table}[hpb]
\scalebox{0.8}{
\begin{tabular}{|c|c|c|c|c|c|c|}
\hline
\multirow{3}{*}{Baselines} & beta       & 0.1             & 0.2              & 0.3              & 0.4              & 0.5              \\ \cline{2-7} 
                           & budget     & 5               & 5                & 5                & 5                & 5                \\ \cline{2-7} 
                           & graph size & 1000            & 1000             & 1000             & 1000             & 1000             \\ \hline
pi                         & IE         & 67.69$\pm$14.2  & 180.92$\pm$28.41 & 355.63$\pm$40.91 & 454.47$\pm$42    & 629.44$\pm$41.11 \\ \hline
degree                     & IE         & 67.11$\pm$14.19 & 180.78$\pm$27.78 & 355.26$\pm$41.52 & 454.66$\pm$46.33 & 629.6$\pm$42.16  \\ \hline
eigen                      & IE         & 63.68$\pm$13.89 & 180.1$\pm$26.34  & 335.37$\pm$38.02 & 514.58$\pm$43.9  & 663.64$\pm$40.43 \\ \hline
RIS                        & IE         & 68.68$\pm$14.36 & 160.48$\pm$26.53 & 343.47$\pm$39.94 & 444.63$\pm$49.08 & 539.68$\pm$46.31 \\ \hline
CELF++                     & IE         & 67.57$\pm$14.31 & 213.98$\pm$33.78 & 356.21$\pm$45.84 & 542.87$\pm$46.09 & 752.41$\pm$40.05 \\ \hline
\end{tabular}
}
\caption{Error Bar of IM Exp 2: Beta}
\label{tab:2}
\end{table}

\begin{table}[hpb]
\scalebox{0.8}{
\begin{tabular}{|c|c|c|c|c|c|c|}
\hline
\multirow{3}{*}{Baselines} & beta       & 0.1             & 0.1             & 0.1             & 0.1             & 0.1             \\ \cline{2-7} 
                           & budget     & 5               & 5               & 5               & 5               & 5               \\ \cline{2-7} 
                           & graph size & 200             & 400             & 600             & 800             & 1000            \\ \hline
pi                         & IE         & 53.17$\pm$10.06 & 57.23$\pm$11.91 & 66.57$\pm$13.52 & 62.54$\pm$13.11 & 68.73$\pm$13.73 \\ \hline
degree                     & IE         & 52.79$\pm$10.27 & 57.58$\pm$11.95 & 66.52$\pm$13.73 & 62.97$\pm$12.72 & 69.33$\pm$13.92 \\ \hline
eigen                      & IE         & 57.81$\pm$10.85 & 59.95$\pm$12.36 & 64.28$\pm$13.65 & 63.73$\pm$13.73 & 66.97$\pm$14.12 \\ \hline
RIS                        & IE         & 56.71$\pm$11.33 & 61.69$\pm$12.38 & 62.96$\pm$14.2  & 61.87$\pm$12.82 & 62.69$\pm$13.53 \\ \hline
CELF++                     & IE         & 57.17$\pm$11.15 & 58.35$\pm$13.37 & 64.62$\pm$14.26 & 62.26$\pm$13.5  & 66.75$\pm$13.64 \\ \hline
\end{tabular}
}
\caption{Error Bar of IM Exp 3: Graph size}
\label{tab:3}
\end{table}

\begin{table}[hpb]
\scalebox{0.8}{
\begin{tabular}{|c|c|c|c|c|c|c|c|}
\hline
Baselines & budget  & 5               & 10              & 15              & 20              & 25              & 30              \\ \hline
greedy    & blocked & 50.65$\pm$19.34 & 91.32$\pm$15.56 & 90.39$\pm$13.65 & 92.12$\pm$17.25 & 93.21$\pm$15.52 & 91.51$\pm$15.22 \\ \hline
eigen     & blocked & 1.56$\pm$22.97  & 1.83$\pm$22.74  & 2.53$\pm$21.68  & -0.4$\pm$22.17  & 5.05$\pm$21.31  & 8.13$\pm$20.98  \\ \hline
degree    & blocked & -1.78$\pm$22.99 & 13.8$\pm$18.89  & 15.75$\pm$19.22 & 16.99$\pm$21.62 & 12.02$\pm$21.46 & 16.59$\pm$22.53 \\ \hline
sigma     & blocked & 1.49$\pm$24.24  & 2.74$\pm$23.99  & 2.95$\pm$21.91  & 5.23$\pm$21.35  & 8.59$\pm$23.14  & 7.73$\pm$21.62  \\ \hline
pi        & blocked & 3.01$\pm$19.62  & 11.36$\pm$18.74 & 10.72$\pm$21.85 & 11.17$\pm$20.18 & 17.47$\pm$23.37 & 17$\pm$18.87    \\ \hline
\end{tabular}
}
\caption{Error Bar of IBM Exp 1: Budget}
\label{tab:4}
\end{table}


\begin{table}[hpb]
\scalebox{0.8}{
\begin{tabular}{|c|c|c|c|c|c|c|}
\hline
Baselines & beta    & 0.1             & 0.2              & 0.3              & 0.4              & 0.5              \\ \hline
greedy    & blocked & 75.33$\pm$12.92 & 194.21$\pm$24.46 & 338.93$\pm$32.47 & 503.26$\pm$40.52 & 653.19$\pm$37.19 \\ \hline
eigen     & blocked & 5.13$\pm$17.69  & 18.63$\pm$32.65  & 32.19$\pm$48.69  & 50.64$\pm$54.89  & 76.48$\pm$55.06  \\ \hline
degree    & blocked & 5.14$\pm$19.98  & 20.48$\pm$30.98  & 48.12$\pm$38.65  & 64.69$\pm$51.74  & 89.5$\pm$56.16   \\ \hline
sigma     & blocked & 3.61$\pm$19.99  & 17.49$\pm$32.88  & 39.89$\pm$48.46  & 62.49$\pm$52.25  & 87.2$\pm$54.2    \\ \hline
pi        & blocked & 5.64$\pm$16.4   & 17.85$\pm$32.51  & 36.35$\pm$39.46  & 67.9$\pm$58.32   & 87.71$\pm$46.58  \\ \hline
\end{tabular}
}
\caption{Error Bar of IBM Exp 2: Beta}
\label{tab:5}
\end{table}



\begin{table}[hpb]
\scalebox{0.8}{
\begin{tabular}{|c|c|c|c|c|c|c|}
\hline
Baselines & graph size & 200             & 400             & 600             & 800             & 1000            \\ \hline
greedy    & blocked    & 38.63$\pm$13.42 & 42.33$\pm$15.63 & 46.15$\pm$18.39 & 46.76$\pm$16.85 & 45.94$\pm$19.24 \\ \hline
eigen     & blocked    & 2.89$\pm$16.95  & 4.6$\pm$23.41   & -0.53$\pm$20.89 & 3.81$\pm$19.07  & -3.12$\pm$22.86 \\ \hline
degree    & blocked    & 1.91$\pm$16.6   & 14.19$\pm$20.63 & 2.16$\pm$18.57  & 4.88$\pm$19.64  & 5.38$\pm$21.07  \\ \hline
sigma     & blocked    & 2.14$\pm$15.74  & 3.75$\pm$17.98  & -0.45$\pm$21.82 & 3.75$\pm$19.57  & 1.42$\pm$18.63  \\ \hline
pi        & blocked    & 5.41$\pm$15.65  & 12.7$\pm$17.95  & 1.84$\pm$18.5   & 6.45$\pm$20.23  & 3.05$\pm$22.25  \\ \hline
\end{tabular}
}
\caption{Error Bar of IBM Exp 3: Graph size}
\label{tab:6}
\end{table}


\begin{table}[hpb]
\scalebox{0.8}{
\begin{tabular}{|c|c|c|c|c|}
\hline
Baselines & Config & CiteSeer & Cora & WS-small world \\ \hline
\multirow{1}{*}{NETSLEUTH (legacy)} & 2 sources & 20.10$\pm$2.13 & 14.50$\pm$3.95 & 11.20$\pm$2.53 \\ \hline
\multirow{2}{*}{NETSLEUTH} & 2 sources & 10.30$\pm$2.45 & 11.10$\pm$2.56 & 11.10$\pm$1.29 \\ \cline{2-5}
 & 3 sources & 5.00$\pm$1.41 & 9.20$\pm$0.84 & 7.60$\pm$1.52 \\ \hline
\multirow{2}{*}{Jordan Centrality} & 2 sources & 11.20$\pm$2.82 & 9.90$\pm$2.64 & 10.80$\pm$1.62 \\ \cline{2-5}
 & 3 sources & 6.60$\pm$5.18 & 7.60$\pm$2.70 & 8.00$\pm$0.00 \\ \hline
\multirow{2}{*}{LISN} & 2 sources & 9.50$\pm$1.84 & 10.20$\pm$4.34 & 10.70$\pm$1.16 \\ \cline{2-5}
 & 3 sources & 3.60$\pm$2.51 & 8.00$\pm$2.83 & 8.00$\pm$1.00 \\ \hline
\end{tabular}
}
\caption{Error Bar of SL Exp: Distance}
\label{tab:7}
\end{table}

\subsection{Runtime}
Table \ref{tab:8}-\ref{tab:14} lists the runtime under various configurations for three tasks described in the main text.

\begin{table}[hpb]
\scalebox{0.8}{
\begin{tabular}{|c|c|c|c|c|c|c|c|}
\hline
\multirow{3}{*}{Baselines} & beta       & 0.1     & 0.1     & 0.1     & 0.1     & 0.1     & 0.1     \\ \cline{2-8} 
                           & budget     & 5       & 10      & 15      & 20      & 25      & 30      \\ \cline{2-8} 
                           & graph size & 1000    & 1000    & 1000    & 1000    & 1000    & 1000    \\ \hline
pi                         & time       & 0.56    & 1.14    & 1.70    & 2.27    & 2.84    & 3.39    \\ \hline
degree                     & time       & 0.01    & 0.01    & 0.01    & 0.01    & 0.01    & 0.01    \\ \hline
eigen                      & time       & 0.09    & 0.17    & 0.38    & 0.34    & 0.41    & 0.64    \\ \hline
RIS                        & time       & 32.71   & 34.57   & 36.07   & 37.98   & 39.22   & 40.47   \\ \hline
CELF++                     & time       & 1166.34 & 2099.72 & 1823.11 & 2363.43 & 2548.16 & 2593.07 \\ \hline
\end{tabular}
}
\caption{Runtime of IM Exp 1: Budget}
\label{tab:8}
\end{table}


\begin{table}[hpb]
\scalebox{0.8}{
\begin{tabular}{|c|c|c|c|c|c|c|}
\hline
\multirow{3}{*}{Baselines} & beta       & 0.1     & 0.2     & 0.3     & 0.4     & 0.5     \\ \cline{2-7} 
                           & budget     & 5       & 5       & 5       & 5       & 5       \\ \cline{2-7} 
                           & graph size & 1000    & 1000    & 1000    & 1000    & 1000    \\ \hline
pi                         & time       & 0.55    & 0.55    & 0.55    & 0.55    & 0.55    \\ \hline
degree                     & time       & 0.01    & 0.01    & 0.01    & 0.01    & 0.01    \\ \hline
eigen                      & time       & 0.09    & 0.09    & 0.09    & 0.09    & 0.09    \\ \hline
RIS                        & time       & 33.15   & 262.63  & 698.52  & 873.41  & 990.61  \\ \hline
CELF++                     & time       & 1089.03 & 1193.37 & 1293.89 & 2180.52 & 2810.78 \\ \hline
\end{tabular}
}
\caption{Runtime of IM Exp 2: Beta}
\label{tab:9}
\end{table}


\begin{table}[hpb]
\scalebox{0.8}{
\begin{tabular}{|c|c|c|c|c|c|c|}
\hline
\multirow{3}{*}{Baselines} & beta       & 0.1   & 0.1    & 0.1    & 0.1    & 0.1     \\ \cline{2-7} 
                           & budget     & 5     & 5      & 5      & 5      & 5       \\ \cline{2-7} 
                           & graph size & 200   & 400    & 600    & 800    & 1000    \\ \hline
pi                         & time       & 0.03  & 0.10   & 0.22   & 0.39   & 0.56    \\ \hline
degree                     & time       & 0.00  & 0.00   & 0.01   & 0.01   & 0.01    \\ \hline
eigen                      & time       & 0.02  & 0.03   & 0.06   & 0.07   & 0.08    \\ \hline
RIS                        & time       & 8.01  & 14.15  & 20.43  & 27.08  & 33.26   \\ \hline
CELF++                     & time       & 78.60 & 332.47 & 498.12 & 801.21 & 1184.42 \\ \hline
\end{tabular}
}
\caption{Runtime of IM Exp 3: Graph size}
\label{tab:10}
\end{table}


\begin{table}[hpb]
\scalebox{0.8}{
\begin{tabular}{|c|c|c|c|c|c|c|c|}
\hline
Baselines & budget & 5       & 10      & 15       & 20       & 25       & 30       \\ \hline
greedy    & time   & 4550.02 & 9062.44 & 13530.73 & 17956.47 & 22369.97 & 26749.04 \\ \hline
eigen     & time   & 0.08    & 0.12    & 0.17     & 0.29     & 0.34     & 0.40     \\ \hline
degree    & time   & 0.01    & 0.01    & 0.01     & 0.01     & 0.01     & 0.01     \\ \hline
sigma     & time   & 0.46    & 0.91    & 1.43     & 1.81     & 2.26     & 2.78     \\ \hline
pi        & time   & 0.45    & 0.88    & 1.30     & 1.71     & 2.15     & 2.55     \\ \hline
\end{tabular}
}
\caption{Runtime of IBM Exp 1: Budget}
\label{tab:11}
\end{table}


\begin{table}[hpb]
\scalebox{0.8}{
\begin{tabular}{|c|c|c|c|c|c|c|}
\hline
Baselines & beta & 0.1      & 0.2      & 0.3      & 0.4      & 0.5      \\ \hline
greedy    & time & 26884.75 & 26932.86 & 26941.75 & 26857.38 & 26857.47 \\ \hline
eigen     & time & 0.41     & 0.42     & 0.41     & 0.42     & 0.42     \\ \hline
degree    & time & 0.01     & 0.01     & 0.01     & 0.01     & 0.01     \\ \hline
sigma     & time & 2.78     & 2.79     & 2.71     & 2.71     & 2.78     \\ \hline
pi        & time & 2.57     & 2.57     & 2.57     & 2.57     & 2.56     \\ \hline
\end{tabular}
}
\caption{Runtime of IBM Exp 2: Beta}
\label{tab:12}
\end{table}


\begin{table}[hpb]
\scalebox{0.8}{
\begin{tabular}{|c|c|c|c|c|c|c|}
\hline
Baselines & graph size & 200    & 400    & 600     & 800     & 1000    \\ \hline
greedy    & time       & 185.96 & 725.73 & 1651.80 & 2945.13 & 4574.89 \\ \hline
eigen     & time       & 0.02   & 0.03   & 0.03    & 0.04    & 0.13    \\ \hline
degree    & time       & 0.00   & 0.00   & 0.00    & 0.01    & 0.01    \\ \hline
sigma     & time       & 0.04   & 0.10   & 0.19    & 0.31    & 0.45    \\ \hline
pi        & time       & 0.02   & 0.08   & 0.16    & 0.28    & 0.45    \\ \hline
\end{tabular}
}
\caption{Runtime of IBM Exp 3: Graph size}
\label{tab:13}
\end{table}


\begin{table}[hpb]
\scalebox{0.8}{
\begin{tabular}{|c|c|c|c|c|}
\hline
Baselines & Config & CiteSeer & Cora & WS-small world \\ \hline
\multirow{1}{*}{NETSLEUTH (legacy)} & 2 sources & 0.89$\pm$0.37 & 1.07$\pm$0.28 & 1.37$\pm$0.55 \\ \hline
\multirow{2}{*}{NETSLEUTH} & 2 sources & 3.71$\pm$2.66 & 32.51$\pm$30.09 & 265.81$\pm$66.08 \\ \cline{2-5}
 & 3 sources & 7.15$\pm$4.21 & 1008.44$\pm$1131.26 & 16227.28$\pm$4704.34 \\ \hline
\multirow{2}{*}{Jordan Centrality} & 2 sources & 4.00$\pm$2.94 & 35.74$\pm$33.46 & 286.32$\pm$64.15 \\ \cline{2-5}
 & 3 sources & 19.00$\pm$29.57 & 1000.00$\pm$1124.15 & 16087.06$\pm$4742.73 \\ \hline
\multirow{2}{*}{LISN} & 2 sources & 71.96$\pm$49.05 & 122.60$\pm$95.47 & 337.84$\pm$66.56 \\ \cline{2-5}
 & 3 sources & 75.05$\pm$74.33 & 1111.07$\pm$1151.89 & 16196.86$\pm$4711.40 \\ \hline
\end{tabular}
}
\caption{Runtime of SL Exp}
\label{tab:14}
\end{table}

\end{document}